\documentclass[review]{elsarticle}

\usepackage{hyperref}
\usepackage{subfigure}
\usepackage{longtable}

\biboptions{authoryear}

\begin{document}
\begin{frontmatter}

\title{Online reviews can predict long-term returns of individual stocks}

\author[mymainaddress]{Junran Wu}
\author[mymainaddress]{Ke Xu}

\author[mysecondaryaddress,mythirdaddress]{Jichang Zhao\corref{mycorrespondingauthor}}
\cortext[mycorrespondingauthor]{Corresponding author:}
\ead{jichang@buaa.edu.cn}

\address[mymainaddress]{State Key Lab of Software Development Environment, Beihang University}
\address[mysecondaryaddress]{School of Economics and Management, Beihang University}
\address[mythirdaddress]{Beijing Advanced Innovation Center for Big Data and Brain Computing}

\begin{abstract}
Online reviews are feedback voluntarily posted by consumers about their consumption experiences. This feedback indicates customer attitudes such as affection, awareness and faith towards a brand or a firm and demonstrates inherent connections with a company's future sales, cash flow and stock pricing. However, the predicting power of online reviews for long-term returns on stocks, especially at the individual level, has received little research attention, making a comprehensive exploration necessary to resolve existing debates. In this paper, which is based exclusively on online reviews, a methodology framework for predicting long-term returns of individual stocks with competent performance is established. Specifically, 6,246 features of 13 categories inferred from more than 18 million product reviews are selected to build the prediction models. With the best classifier selected from cross-validation tests, a satisfactory increase in accuracy, 13.94\%, was achieved compared to the cutting-edge solution with 10 technical indicators being features, representing an 18.28\% improvement relative to the random value. The robustness of our model is further evaluated and testified in realistic scenarios. It is thus confirmed for the first time that long-term returns of individual stocks can be predicted by online reviews. This study provides new opportunities for investors with respect to long-term investments in individual stocks.
\end{abstract}

\begin{keyword}
\texttt online reviews \sep predicting model \sep long-term returns \sep individual stocks
\end{keyword}

\end{frontmatter}


\section{Introduction}
As Adam Smith wrote in The Wealth of Nations, everyone aims to use his or her capital to gain the most value from his or her products \citep{smith1937wealth}, which tells us that pursuing personal interests is the dominant motivation for human beings to engage in economic activities. Moreover, it is evident that those who invest in the stock market expect excess returns. Owing to the profitable nature of stocks, investors have expended considerable effort on stock prediction both in theoretical and application perspectives. Facilitated with interior information such as historical prices and external signals such as investor behaviors, predicting returns of stocks can help guide trading decisions in asset markets. Additionally, with the information explosion, data related to the financial market, both in source and volume, have been enriched and gradually accumulated. Price information at various frequencies \citep{harris1986transaction, jain1988dependence, pan2017multiple}, companies’ financial reports \citep{jones1970quarterly, zhou2015performance, zhou2017predicting}, and financial news \citep{geva2014empirical, li2014effect, nassirtoussi2015text} are examples of direct information related to the financial system. In the meantime, the rise and fall of macroeconomics \citep{chen1986economic}, as well as the reactions and reflections from investors’ emotions unveiled by social media \citep{zhou2018tales, sun2017predicting, ruan2018using, zhang2018improving}, search engines \citep{preis2013quantifying, engelberg2011search, xu2019does} and analysts’ recommendations \citep{duan2013posterior}, are indirect yet inspiring sources of data.

Of all these various data sources, customer-generated online reviews are among the newest. Online reviews are voluntarily posted on e-commerce websites by customers looking to share their consumption experiences \citep{clemons2006online, godes2009firm, tang2012content}. Jeff Bezos, the CEO of Amazon, described the power of online customer reviews as “If you make customers unhappy in the physical world, they might each tell 6 friends. If you make customers unhappy on the internet, they can each tell 6,000 friends.” As user-generated word-of-mouth messages, consumer reviews not only provide signals about a company’s products but also affect consumers’ decisions \citep{cabral2000stretching}. From a business perspective, reviews indicate customer attitudes, such as affection, awareness and faith, towards a brand or firm \citep{chevalier2006effect, li2008self}, and brand image is in fact inherently related to firm equity \citep{faircloth2001effect}. As the target of a company’s products and services, customers not only actively produce product-related information that is readily available to other consumers but also regularly adapt such information when making purchase decisions, which certainly impacts a company’s future sales and profitability \citep{subrahmanyam1999going, huang2018customer}. \citet{luo2009quantifying} even revealed that online reviews might impact the cash liquid of companies and eventually reshape the fluctuations of their stocks. In addition, a significant predictive relationship between online consumer reviews and firm equity value has been revealed, e.g., customer decisions drive a firm’s equity value \citep{luo2013social}. More importantly, reviews are often related to one specific brand and are dated with the time they are first posted, thus making it possible for data to be collected in individual stock granularity and to track consumer opinions over long periods. Intuitively, these data can be directly connected with the long-term returns of individuals stocks and therefore serve as a new but powerful supplement to the existing data sources in stock prediction practices. However, whether online reviews can be helpful in long-term return prediction continues to lack a comprehensive understanding, especially at the level of individual stocks.

Stock returns can be predicted in different granularities, such as long-term (weekly or monthly) or short-term (in minutes or days). While short-term prediction always aims to capture the instant fluctuations and offer immediate investment advice, long-term prediction helps in value investing, which places greater emphasis on the low frequency trading patterns targeting long-term returns. In the early stock prediction literature, much greater effort was devoted on daily price changes rather than long-term price changes \citep{jasemi2011modern, ye2016novel, baralis2017discovering}. However, for both human and algorithms, long-term prediction is usually more challenging than the short-term counterpart~\citep{ding2014using, zhang2017stock}, which urgently implies the need to build more competent models to predict long-term returns. Meanwhile, instead of rare attempts to predict returns of individuals stocks, existing studies have focused more on predicting the entire market, such as the index \citep{hsu2011hybrid, chen2015hybrid, efendi2018new}. The significance of relationships
between online reviews and stocks, the most common concern in existing studies, has been extensively investigated; however, using reviews as features to directly predict stocks has been rarely explored, and a systematic understanding is still missing. Given the fine level of granularity in decision-making, individual stock prediction might be more powerful and practical for investors in making investments. Because of the long duration and relevance to individual brands, online reviews are actually capable of providing an unprecedented opportunity to address the shortage of existing studies in long-term individual stock forecasting. It is also worth noting that the preliminary explorations on connections between reviews and stock performance mostly utilize only one or two attributes of reviews, such as the rating or the volume \citep{huang2018customer, chen2012third, luo2013consumer, luo2013social}, with other information that might deliver unexpected predictive power being neglected. Those deliberately ignored attributes, such as the time difference between order date and review date, sellers' replies to reviews and devices consumers used to post reviews, will be promising features for predicting long-term returns at the individual level.

In this paper, which is based exclusively on online reviews, a novel framework for predicting long-term returns of individual stocks is established with competent performance and practical implications. Specifically, from JD.com, the largest self-operated\footnote{Most products sold by JD} online retailer in China, more than 18 million online product reviews of 102 firms are thoroughly collected. In total, based on our new feature extraction methods, over 7,000 features of 13 categories are accordingly derived and inferred. With completion of Pearson correlation analysis, we choose the eight-week return as our long-term prediction target. Through 5-fold cross-validations in 14,688 firm-week samples by gradient boosting algorithm, 6,246 features have been confidently selected to remove noise and achieve relatively high accuracy. Based on these features, a comprehensive comparison of various classifier algorithms, including both state-of-the-art solutions and classical approaches, demonstrates that XGB (eXtreme Gradient Boosting) is the best model with 59.65\% average accuracy in 5-fold cross-validation. This model also achieves 61.02\% accuracy in the hold-out test and greatly exceeds the model with financial technical indicators being features as high as 13.94\%; this implies its strong potential in realistic applications. More surprisingly, our model is also robust to the cut-off points of categorizing returns and shows even more power in predicting abnormal cases. It is also interesting that to obtain the most satisfying accuracy, the length of training window is nearly three years, which is theoretically consistent with the time required for the formation of a brand image.

The remainder of this paper is organized as follows. Section 2 discusses the related literature. Section 3 introduces the online review data and the stock price data. Features and targets are depicted in Section 4, and Pearson correlation analysis and target selection are also conducted in this section. Section 5 describes how to select features and build classification models, and the hold-out test is also conducted to evaluate the performance in realistic investments. The robustness of our model is further tested in Section 6, and a brief conclusion is provided in Section 7.

\section{Related Work}
Because of the decisive role of stock trend prediction in investments, researchers have devoted many efforts to study it. Before the information explosion caused by the internet, technical analysis was the main approach for trend prediction. The autoregressive (AR) model has been one of the most widely used for stationary and linear time-series \citep{li2016stock}. To overcome the nonstationary and nonlinear nature of stock prices, nonlinear learning models have been presented to catch the intricate pattern hidden in market trends \citep{nayak2015naive, pan2017multiple}. After the recent rise of neural networks, more research efforts have been allocated to exploiting deep learning models in financial prediction \citep{long2019deep, goccken2016integrating, kim2012simultaneous, patel2015predicting, o2006neural}.

One major limitation of technical analysis is that it is not able to uncover the principles that rule the dynamics of the market through price changes alone. Researchers have therefore sought information outside the market to improve prediction performance. Substantial online content comes with the information explosion, such as companies’ financial reports \citep{jones1970quarterly, zhou2015performance, zhou2017predicting}, financial news \citep{geva2014empirical, li2014effect, nassirtoussi2015text}, signals from the rise and fall of macroeconomics \citep{chen1986economic}, the reactions and reflections from investors’ emotions delivered in social media posts \citep{zhou2018tales, sun2017predicting, ruan2018using, zhang2018improving, li2014effect}, search engines queries \citep{preis2013quantifying, engelberg2011search, xu2019does} or analysts’ recommendations \citep{duan2013posterior}. A new perspective to understand the stock market more comprehensively and make return predictions more accurate has come to light. Many attempts have been made to mine internet data for better market trend prediction. \citet{ding2015deep} focused on event-driven stock market prediction through a deep learning model. \citet{wang2014semiparametric} proposed a text regression model to predict the volatility of stock prices. \citet{hagenau2013automated} extracted a wide range of features to represent unstructured text data and performed a robust feature selection on stock prediction to improve accuracy. \citet{zhou2016can} concentrated on the Chinese stock market, assigning five kinds of emotions to more than 10 million stock-relevant tweets, showing that part of these emotions could be used in predicting the Chinese stock market index.~\citet{wang2019aggregating} even established a prediction-independent framework to fuse data of various types.

However, these early stock prediction studies are more concerned with daily rather than long-term price changes \citep{jasemi2011modern, chen2015hybrid, baralis2017discovering} because for both humans and algorithms, the performance of short-term prediction is usually better than long-term prediction \citep{ding2014using, zhang2017stock}. Meanwhile, existing studies focus more on the prediction of the entire market, such as with indices \citep{hsu2011hybrid, chen2015hybrid, efendi2018new}, and have rarely made return predictions for individuals stocks, which is more practical and direct in realistic investments, especially considering the vast number of inexperienced and emotional individual investors in China~\citep{zhou2018tales}.

Online reviews, one new form of Internet content that reflects customer attitudes towards products and firms \citep{chevalier2006effect, li2008self}, have been proven to have connections with companies’ future sales and profitability \citep{subrahmanyam1999going, huang2018customer}. \citet{chevalier2006effect} found that firm sales are positively correlated with consumer review ratings, and \citet{hu2009identifying} showed that online review ratings lower consumer uncertainty, though some customers prefer some uncertainty in their consumption experience\citep{martin2007choosing}. \citet{liu2006word} found that movie reviews impact box office revenues. \citet{senecal2004influence} showed that consumers who refer to online reviews in shopping decisions are more likely to select recommended products. \citet{morgan2006value} suggested that firms with a good reputation among customers are more likely to experience growth in firm equity. \citet{tellis2007value} revealed that the valuation of firm products given by investors would be affected by review ratings on product quality. Due to the long duration and the inherent relevance to brands, online reviews therefore indeed offer a new opportunity to overcome the difficulties that existing studies have experienced regarding long-term individual stock forecasting.

The prediction ability of online product reviews for stock returns \citep{tirunillai2012does, chen2012third, luo2013consumer, luo2013social}, in particular for the short-term price changes, has been preliminarily examined. \citet{luo2013social} suggested that online consumer reviews have a significant predictive relationship with firm equity value. Additionally, \citet{tirunillai2012does} found a complicated result by investigating the relationship among product reviews and stock market variables. However, the firms that these studies generally examine are a relatively small set. There is obviously one exception: \citet{fornell2016stock} obtained a sample of approximately 300 firms over 15 years using annual customer satisfaction scores and used them to construct an investment strategy, which recorded 518\% cumulative returns over the 15 years. Similar to this study, \citet{huang2018customer} found evidence that online consumer reviews contain novel information for stock pricing; moreover, a spread portfolio that is long on stocks with high abnormal customer ratings and short on stocks with low abnormal customer ratings delivers an abnormal return of approximately 55.7 to 73.0 basis points per month.

Nevertheless, few of these landmark efforts to explore the connection between online reviews and stock pricing, which inspired the present study, paid sufficient attention to the predictive ability of online reviews. A comprehensive investigation and a clear conclusion on whether online reviews can be used for long-term stock prediction is still missing. A niche solution for return predictions of individual stocks is accordingly lacking. This paper complements these studies by highlighting product-related reviews of consumers as an important source of information that could be used for long-term stock return prediction on an individual level.

\section{Data}
In this section, details of how customer-oriented companies and their online product reviews and stock prices are retrieved and collected will be illustrated.

\subsection{Customer-oriented companies}
To ensure the availability of consumer reviews, we first had to collect a list of companies in the customer market. Sina Finance is the largest financial news portal in China. Since its inception in August 1999, this website has continuously provided news and information in the financial industry and accounts for more than one-third of the financial websites market. Based on the characteristics of stocks, Sina Finance divides the whole stock market into multiple sectors. In accordance with the industry classification principle, the home appliance industry, garment industry, wine industry and food industry represent the four sectors most relevant to consumer products after we retrieved all the sectors and stocks of each sector presented on Sina Finance.

In these four sectors, there are 177 firms with a market value ranging from the billions to hundreds of billions, covering all kinds of market capitalization companies. We noted the names and codes of these stocks as seeds for both online review crawling and stock price data acquisition.

\subsection{Online reviews from JD.com}
JD is one of the two largest B2C\footnote{Business-to-Customer} online retailers in China, generating \$193 billion gross merchandise volume in 2017. With the completion of the 3C\footnote{Computer, Communication and ConsumerElectronics} product line in 2008, customers began generating reviews for the products they bought. Since then, more than 300 million customers have purchased on JD.com, along with more than one billion reviews have been posed. According to JD’s review creation guidelines, by providing references for other consumers about shopping decisions and business decision-making, consumers can make a fair, objective and true evaluation of the order after the transaction is completed.\footnote{\url{https://rule.jd.com/rule/ruleDetail.action?ruleId=2395}} Reviews are made up of text and a rating on a scale from one to five stars, with five being the top rating. All reviews are dated by the time they are first posted, which makes it possible to track consumer opinions over time. These reviews regard not just the product itself but also everything about the seller, including the shipping and delivery experience, as well as anything that reflects a customer’s impression of the brand that may affect investor decisions \citep{chevalier2006effect, li2008self}. Additionally, JD is the largest self-operated e-commerce platform in China, which means that JD sells most products on the platform itself; thus, there is no conflict of interest, no paid reviews or sellers posting positive reviews for their own products or negative reviews of competing products \citep{malbon2013taking, zhang2016online}. Sophisticated technical tricks such as captcha are also employed to prevent possible bots and spam. All these aspects ensure that the product reviews on JD are of a high quality and reliable.

To identify public firms with customer product reviews on JD within the 177 seeds, we first checked whether the firm sells its products on JD.com by manually searching for the company name. We then retrieved the list of brands from JD.com under each product category and identified the companies that own these brands within the seeds. In all, we obtained 109 public firms that have customer product reviews on JD.com.

\begin{table}[!htbp]
\centering
\setlength{\belowcaptionskip}{10pt} 
\caption{Summary statistics on JD.com reviews for 102 public firms}
\label{tb_review}
\resizebox{\textwidth}{!}{
\begin{tabular}{lccc}
\hline
 & \multicolumn{1}{l}{Number of reviews} & \multicolumn{1}{l}{Number of products} & \multicolumn{1}{l}{Number of firms} \\ \hline
Final Sample & 18,008,415 & 164,715 & 102 \\
Home Appliance Industry & 7,245,982 & 44,059 & 25 \\
Garment Industry & 5,365,380 & 75,774 & 27 \\
Food Industry & 4,021,617 & 22,964 & 29 \\
Wine Industry & 1,375,436 & 21,918 & 21 \\ \hline
\end{tabular}
}
\end{table}

To collect the reviews for the sample of public firms, we developed a web-crawling program that receives a brand's name owned by a public firm as a search term on JD.com and generates a list of all products whose brand name perfectly matches the search term. For each product, we retrieved all the reviews associated with it through the review interface by traversing all the stars and pages. From the review interface, in addition to the numerical star rating, the date and the text of one review, we also obtain the days between the order date and the review date, the number of useless and useful votes, comments from the sellers on this review, the image list attached to it, the client by which the review was produced (e.g., iPhone, iPad, Android or Web), etc. A snippet of reviews collected can be found in Figure \ref{review_sample}. The sample of the reviews that we crawled covers the period from November 2008 to December 2017, and we remove duplicate reviews posted with the same review ID for the same product.\footnote{Because many products link to the same page, many reviews will be crawled repeatedly from the same product page.} To reduce noise in the sample data, we require that a firm have at least 1,000 online reviews and that the timespan of a firm’s review should be longer than twelve months. Table \ref{tb_review} reports the number of reviews, products and firms for the final sample as well as by the four industry sectors. Table \ref{tb_firm_review} lists the detail about firms used in this paper. More than 18 million reviews on 164,715 products manufactured by the sample firms are posted. The top two industries in terms of the number of product reviews are the home appliance industry (7.2 million reviews) and garment industry (5.3 million reviews), accounting for 70\% of the reviews that we acquired.

After sorting them by the review date, we merged all firms’ product reviews into one time series. As shown in Figure \ref{comment_number}, we used a half year as the time window for the presentation; before 2014, a relatively small number of reviews were posted on JD.com in our four seed sectors. This limited number is because the internet was not yet universal in China and people were not familiar with online shopping during that period.\footnote{The 35th Statistical Report on Internet Development in China from CNNIC, \url{http://www.cac.gov.cn/2015-02/03/c_1114222357.htm}} Around 2014, however, given the ubiquity of the mobile Internet industry, there has been exponential growth in online shopping, which has caused product reviews to accumulate rapidly.

\subsection{Stock price data}
The price of one stock is typically affected by supply and demand of market participants. However, some corporate actions also affect a stock's price, which needs to be adjusted after these actions, such as stock splits, dividends or distributions and rights offerings. The adjusted price is a useful tool when examining historical returns because it delivers an accurate representation of a firm's equity value beyond the simple market price. We obtained the adjusted daily price data from Tushare\footnote{\url{http://tushare.org}}, a free, open source Python financial data interface package supported by Sina Finance, Tencent Finance, Shanghai Stock Exchange and Shenzhen Stock Exchange.

\begin{table}[!htbp] 
\centering
\setlength{\belowcaptionskip}{10pt} 
\caption{Basic Review Features}
\label{tb_features_basic}
\resizebox{\textwidth}{!}{
\begin{tabular}{ll} \hline
Feature & Definition \\ \hline
$W_nReview$ & $ N $ \\ 
$W_nStar^s$ & $ \sum_{i=1}^N\{1\ \mbox{if}\ star_i==s\ \mbox{else}\ 0\} $ \\
$W_nStar^{15}Diff$ & $ W_nStar^5_i - W_nStar^1_i $ \\
$W_nDefault$ & $ \sum_{i=1}^N\{1\ \mbox{if}\ isDefault_i==True\ \mbox{else}\ 0 \}$ \\
$W_nScore$ & $ \frac{\sum_{s=1}^5(s\times W_nStar^s_i)}{W_nReview_i} $ \\ 
$W_nEmotion^e$ & $ \sum_{i=1}^N\{1\ \mbox{if}\ ReviewEmotion_i==e\ \mbox{else}\ 0\} $ \\ 
$W_nEmotion$ & $ \sum_{e=0}^4{W_nEmotion^e} $ \\
$W_nEmotion^{negative}$ & $ \sum_{i=1}^N\{1\ \mbox{if}\ ReviewEmotion_i\in\{0, 1, 3, 4\}\ \mbox{else}\ 0\} $ \\
$W_nTendency^{posW}$ & $ \sum_{i=1}^NReviewPos_i $ \\ 
$W_nTendency^{negW}$ & $ \sum_{i=1}^NReviewNeg_i $ \\
$W_nTendency^{word}$ & $ \frac{W_nTendency_{pos}-W_nTendency^{neg}}{W_nTendency_{pos}+W_nTendency^{neg}} $ \\
$W_nTendency^{posR}$ & $ \sum_{i=1}^N\{1\ \mbox{if}\ ReviewPos_i>ReviewNeg_i\ \mbox{else}\ 0\} $ \\ 
$W_nTendency^{negR}$ & $ \sum_{i=1}^N\{1\ \mbox{if}\ ReviewPos_i<ReviewNeg_i\ \mbox{else}\ 0\} $ \\ 
$W_nTendency^{pos}$ & $ \sum_{i=1}^N\{ReviewTen_i \ \mbox{if}\ ReviewPos_i>ReviewNeg_i\ \mbox{else}\ 0\} $ \\
$W_nTendency^{neg}$ & $ \sum_{i=1}^N\{ReviewTen_i \ \mbox{if}\ ReviewPos_i<ReviewNeg_i\ \mbox{else}\ 0\} $ \\
$W_nTendency$ & $ W_nTendency^{pos} + W_nTendency^{neg} $ \\
$W_nDays$ & $ \frac{\sum_{i=1}^Ndays_i}{W_nReview}$ \\
$W_nUseful$ & $ \sum_{i=1}^NusefulVoteCount_i$ \\
$W_nUsefulR$ & $ \sum_{i=1}^N\{1\ \mbox{if}\ usefulVoteCount_i>0\ \mbox{else}\ 0\}$ \\
$W_nUseless$ & $ \sum_{i=1}^NUselessVoteCount_i$ \\
$W_nUselessR$ & $ \sum_{i=1}^N\{1\ \mbox{if}\ UselessVoteCount_i>0\ \mbox{else}\ 0\}$ \\
$W_nImage$ & $ \sum_{i=1}^Nimg_i$ \\
$W_nImageR$ & $ \sum_{i=1}^N\{1\ \mbox{if}\ img_i>0\ \mbox{else}\ 0\}$ \\
$W_nReply$ & $ \sum_{i=1}^NreplyCount_i$ \\
$W_nReplyR$ & $ \sum_{i=1}^N\{1\ \mbox{if}\ replyCount_i>0\ \mbox{else}\ 0\}$ \\
$W_nClient^c$ & $ \sum_{i=1}^N\{1\ \mbox{if}\ userClient_i==c\ \mbox{else}\ 0 \}$ \\
$W_nMobile$ & $ \sum_{i=1}^N\{1\ \mbox{if}\ isMobile_i==True\ \mbox{else}\ 0\}$ \\ \hline
\end{tabular}}
\end{table}

\section{Features \& Targets}
In this section, 13 categories of features extracted from online reviews and 12 types of weekly stock returns as long-term targets are depicted. Through Pearson correlation analysis, the best target to describe the relationship between online reviews and firm equity value will also be determined.

\subsection{Features}
In addition to the known features, which have a significant relationship with firm equity, we extracted other features reflecting the consumer experiences and consumer images. To aggregate review data for feature extraction, we transform the single review to firm-weeks data. For each firm, all the reviews from different products are integrated by time. How 13 categories of basic features are extracted from firm-weeks data for each firm is listed in Table~\ref{tb_features_basic}, and all the features have one time point per week. Note that in Table~\ref{tb_features_basic}, $N$ is the number of reviews in $n$ weeks and $n\in[1, 12]$; $M$ is the number of reviews in $[i-n-m, i-n]$ weeks, where $m$ is the length of relative history data window and $m \in \{4,6,8,10,12,16,20,24|m>n\}$. Apart from the basic features, we also make various transformations from the basic features to enrich signals derived from reviews, which are shown in Table \ref{tb_features_variant}. For the meaning of each basic feature, each category is described in detail in its own section.

In early studies, researchers studied several attributes of online reviews, such as volume, semantic polarity, rating and emotion. Furthermore, a significant relationship between online reviews and firm equity and the surprise profitability of abnormal rating are also revealed from these features. Referring to these early efforts, the six categories of features that we extracted are listed below.

\begin{enumerate}[(1)]
\item \textbf{Review}. The $Review$ category is constituted by features describing the volume of reviews. \citet{chen2012third, luo2013social} have shown a significant relationship between the online information volume and firm equity. The basic feature in this category is $W_nReview$, which represents the number of reviews in $n$ weeks. All features related to $Review$ are listed in Appendix Table \ref{tb_features_review}.

\item \textbf{Star}. The $Star$ category contains features extracted from the review rating, which indicates customer satisfaction with the purchase experience. \citet{huang2018customer, luo2013consumer, luo2013social} have demonstrated the power of the review rating on stock earnings. The basic features in this category are $W_nStar^s$, where $s \in [1, 5]$ and $W_nStar^{15}$, representing the number difference between the top 5 rating reviews and the rating reviews below 1 in $n$ weeks. All features related to $Star$ are listed in Appendix Table \ref{tb_features_star}.

\item \textbf{Default}. The $Default$ category contains features extracted according to whether a review is generated by the system automatically due to the comment window closing, which is denoted by $isDefault$. The basic feature in this category is $W_nDefault$, which represents the number of default comments with top 5-star rating in $n$ weeks. All features related to $Default$ are listed in Appendix Table \ref{tb_features_default}.

\item \textbf{Score}. The $Score$ category is composed of features based on the average rating in several weeks, which reflect average customer satisfaction with all purchases in $n$ weeks. \citet{huang2018customer} showed a spread portfolio that relies on score changes and delivers an abnormal return. The basic feature in this category is $W_nScore$. All features related to $Score$ are listed in Appendix Table \ref{tb_features_score}.

\item \textbf{Emotion}. The $Emotion$ category includes features extracted from the emotions delivered by each review, which represent the specific attitudes towards the purchase experience. Recent studies have successfully incorporated investors’ emotions from social media into predicting stock prices \citep{wang2019aggregating, zhou2018tales, sun2017predicting, ruan2018using, li2014effect}. In line with these studies, the emotion measures from \citep{zhou2018tales} are employed here to arrange short texts of reviews into five categories of “anger”, “disgust”, “joy”, “sadness”, and “fear”, which are denoted as $0,\; 1,\; 2,\; 3,\; \textrm{and} \; 4$, respectively. The emotion of each review is denoted by $ReviewEmotion$. The basic features in this category are $W_nEmotion^e$, $W_nEmotion$ and $W_nEmotion^{negative}$, separately the number of reviews possessing emotion $e$ in $n$ weeks, the number of emotional reviews, and the number of reviews that deliver negative emotions including “anger", “disgust", “sadness" and “fear". Note that the emphasis on negative emotions comes from the evidence demonstrated in~\citep{luo2009quantifying, bambauer2011brand} that negative reviews can undermine a brand image in a way that lasts months. In addition, special variants of the category of $Emotion$, i.e., $RatioE$, $RatioEDiff$ and $RatioEDiffH_m$, are further illustrated in Table \ref{tb_features_emotionVariant}. All features related to $Emotion$ are listed in Appendix Table \ref{tb_features_emotion}.

\item \textbf{Tendency}. The $Tendency$ category is composed of features extracted according to tendentious words in a review text. The different tendency reviews have a different relation with stock return \citep{tirunillai2012does, luo2009quantifying}. The number of positive words in a review is denoted by $ReviewPos$, and the number of negative words in a review is denoted by $ReviewNeg$. The tendency of a review is denoted by $ReviewTen=\frac{ReveiwPos - ReviewNeg}{ReviewPos + ReveiwNeg}$ when there is at least one tendentious word in this review; otherwise, it is $0$. The basic features in this category are $W_nTendency^{posW}$, $W_nTendency^{negW}$, $W_nTendency^{word}$, $W_nTendency^{posR}$, $W_nTendency^{negR}$, $W_nTendency^{pos}$, $W_nTendency^{neg}$ and $W_nTendency$, respectively, the number of positive words in all reviews within $n$ weeks, the number of negative words in all reviews within $n$ weeks, the phased tendency at the word level, the number of reviews that contain more positive tendency words than negative ones, the number of reviews that contain more negative tendency words than positive ones, the cumulative tendency for positive reviews, the cumulative tendency for negative reviews and the entire tendency within $n$ weeks. All features related to $Tendency$ are listed in Appendix Table \ref{tb_features_tendency}.
\end{enumerate}

Previous studies of consumers and marketing have shown that experiences occur when consumers search for, shop for, examine, evaluate and consume products \citep{arnould1993river, holbrook2000millennial, hoch2002product}. Both \citet{huffman1993goal} and~\citet{hoch1989managing} have revealed the influence of product experiences on product judgments, attitudes, preferences, purchase intent, and recall. Moreover, customers interact with salespeople and that experience affects customers’ feelings, brand attitudes, and satisfaction \citep{grace2004examining}. Given the importance of consumer experience in brand attitude formation yet few discussions about the relationship between consumer experience and firm equity, five new categories of features extracted from online reviews as a proxy of consumer experience are listed below.

\begin{enumerate}[(1)]
\item \textbf{Days}. The $Days$ category is composed by features extracted from the $days$ field, which describes the date difference from order date to comment date of the corresponding product. This date difference consists of delivery time and part of experience time, which is a type of product experience. The basic feature in this category is $W_nDays$, which represents the average $days$ for all reviews in $n$ weeks. There is one special variant $W_nDaysH_m = W_nDays_{i}\times W_nReview_i - W_mDays_{i-n}\times W_mReview_{i-n}$ that shows the difference between the total days value in $n$ weeks and that in the prior $m$ weeks. All features related to $Days$ are listed in Appendix Table \ref{tb_features_days}.

\item \textbf{Useful}. The $Useful$ category includes features extracted from the $usefulVoteCount$ field which denotes the number of consumers who thought the review is useful. Useful votes in online reviews give advice for other consumers to make purchase decision when examining and evaluating products. The basic features in this category are $W_nUseful$ and $W_nUsefulR$, respectively, the cumulative value of $usefulVote$ for all reviews and the number of reviews that earned at least one useful vote in $n$ weeks. All features related to $Useful$ are listed in Appendix Table \ref{tb_features_useful}.

\item \textbf{Useless}. The $Useless$ category consists of features extracted from the $uselessVoteCount$ field, which contrarily indicates the number of customers who thought that a review is useless. The basic features in this category are $W_nUseless$ and $W_nUselessR$, the cumulative value of $uselessVote$ for reviews and the number of reviews receiving at least one useless vote in $n$ weeks, respectively. All features related to this category are listed in Appendix Table \ref{tb_features_useless}.

\item \textbf{Image}. The $Image$ category is composed of features extracted from the $image$ field, which lists the images that customers post with the review. The images listed in reviews are more vivid and objective than the official images, which helps consumers with product selection. The number of images in a review is denoted by $img$. The basic features in this category are $W_nImage$ and $W_nImageR$, which represent the total number of images and the number of reviews with images in $n$ weeks, respectively. All features related to $Image$ are listed in Appendix Table \ref{tb_features_image}.

\item \textbf{Reply}. The $Reply$ category includes features extracted from the $replyCount$ field, which represents the number of replies from the official sellers as consumer interactions with salespeople. The basic features in this category are $W_nReply$ and $W_nReplyR$, which represent the total number of replies and the number of reviews with attached replies in $n$ weeks, respectively. All features related to $Reply$ are listed in Appendix Table \ref{tb_features_reply}.
\end{enumerate}

Apart from known features and consumer features, we also extracted two categories of features representing consumer images listed below, which are ignored in previous exploitations. \citet{parker2009comparison} has shown that consumer images are related to brand image; furthermore, the image a person has of herself/himself often influences the brands that individuals purchase \citep{plummer2000personality, belk1988possessions, sirgy1982self}. As the proxy of consumer image, the devices that consumers use to post reviews are employed. Except for client, whether the client is a mobile device is also a factor that affects purchase intention \citep{wang2016mobile, holmes2013mobile}.

\begin{enumerate}[(1)]
\item \textbf{Client}. The $Client$ category is composed of features extracted from the $userClient$ field, which reflects the device on which consumers post reviews. There are four kinds of devices: “Web", “iPhone", “Android" and “WeChat", which are denoted by $0,\; 2,\; 4,\; \textrm{and} \; 21$, respectively. The basic features in this category are $W_nClient^c$, which reflects the number of reviews posted from a given client $c$. All features related to $Client$ are listed in Appendix Table \ref{tb_features_client}.

\item \textbf{Mobile}. The $Mobile$ category contains features extracted from the $isMobile$ field, which shows whether the users comment on the product through a mobile device or not. The basic feature in this category is $W_nMobile$, which is the number of reviews posted through mobile devices within $n$ weeks. All features related to $Mobile$ are listed in Appendix Table \ref{tb_features_mobile}
\end{enumerate}

Through the above methods, we extract 7,960 features in total from various attributes of online product reviews. In addition to the known 6 categories of features that have been proven useful for stock pricing and earning in early studies, the other 7 categories of features that represent consumer experiences and images are also extracted to take full advantage of information carried in online reviews. Furthermore, in addition to using only 13 categories of basic features, we calculate various differential sequences from basic features not only for stationarity but also to reflect the change in customer attitudes towards brands. In all, we provide one novel set of feature extraction methods for future studies in stock return prediction.

\subsection{Targets}
Contrary to extensive efforts devoted to daily stock returns, existing studies have rarely examined weekly stock returns and their predictions~\citep{hsu2011hybrid, jasemi2011modern, chen2015hybrid}. This vital gap of missing long-term prediction models motivates the present study to investigate weekly returns of individual stocks. As shown in Equation (\ref{weekly_price}), we transformed the daily stock price data to weekly price data, which is in line with the feature's granularity. The $1$ and $-1$ in Equation (\ref{weekly_price}) stand for the first and the last trading day in the week, respectively.

\begin{equation}
\begin{array}{ll}
  CLOSE_{W_{i}} &= CLOSE^{W_i}_{TD_{-1}} \\
  OPEN_{W_{i}} &= OPEN^{W_i}_{TD_{1}} \\
  LOW_{W_{i}} &= min\{LOW^{W_i}_{TD_{j}}|j\in[1, 5]\} \\
  HIGH_{W_{i}} &= max\{HIGH^{W_i}_{TD_{j}}|j\in[1, 5]\}  \\
\end{array}
\label{weekly_price}
\end{equation}

The weekly stock return is defined in Equation. ({\ref{weekly_stock_return}}) as
\begin{equation}
 RW_n = \frac{CLOSE_{W_{n+i}}-CLOSE_{W_n}}{CLOSE_{W_n}}, \qquad i \in [1, 12] \label{weekly_stock_return}
\end{equation}
ranging from one to twelve weeks.

\subsection{Target Selection}
In the above sections, we describe the features and targets, which will contribute to revealing the correlation between online reviews and stock returns. However, the purpose of this paper is to find out whether online reviews can predict long-term, e.g., weekly stock returns. Moreover, because of the long lag of online information impact \citep{luo2009quantifying, tirunillai2012does}, there is a shift of the online review features to an earlier date, ranging from 1 to 12 weeks. Hence, each feature corresponds to 12 time series according to the shifted time.

Before the analysis of the relation between feature and target, we first normalize all the time series in which data items are transformed to the values from 0 to 1 as $T_{i} = (T_{i} - T_{min})/(T_{max} - T_{min}),$ $T$ represents an arbitrary time series of feature or target, $T_{i}$ is the $i$-th item in time series $T$, and $T_{max}$ and $T_{min}$ are the maximal and minimal value of $T$, respectively. Then, through Pearson correlation analysis, we measure the linear relationship between features and targets as $\rho.$

To observe which target is the best choice for weekly stock return predictions, we need to determine which targets have the best correlation with all the features in 102 firms. Moreover, except for $\rho$, the $p$-value, defined as $\frac{\rho\sqrt{n-2}}{\sqrt{1-\rho^2}}$, where $n$ is number of observations, is also employed to reflect the significance of the correlation between feature and target. A small $p$-value is an indication that the null hypothesis is false, allowing us to conclude that the correlation coefficient is different from zero and that a linear relationship exists. In general, 0.05 is the critical value in determining whether the correlation is significant. In our study, the $p$-value of a feature $f$ and a target $t$ in a specific firm $i$ with a shifted time $s$ is denoted as $PValue_{fsti}$.

For target selection, we count the significant number of firms for every feature-target pair with 0.05 as the critical p-value. Considering that every feature corresponds to 12 times series, we first check the significance of one feature-target pair for a specific firm, which is calculated as Equation (\ref{pearson_count_shift}). When there is one $p$-value in twelve shifted time series that is less than 0.05, we confirm that this feature-target pair in firm $i$ is significant and assign the corresponding $Significance_{fti}$ to 1. The count formula is shown in Equation (\ref{pearson_count}), where $I$ if the number of sample firms.

\begin{equation} 
Significance_{fti}=\left\{
\begin{array}{ll}
1, &\exists s\ PValue_{fsti} \leq 0.05, s\in[1, 12]\\
0, &otherwise.
\end{array}
\right.
\label{pearson_count_shift}
\end{equation}

\begin{equation}
Count_{ft} = \sum_{i=1}^I{Significance_{fti}}
\label{pearson_count}
\end{equation}

\begin{figure}[!htbp]
  \centering
  \includegraphics[width=1.\textwidth]{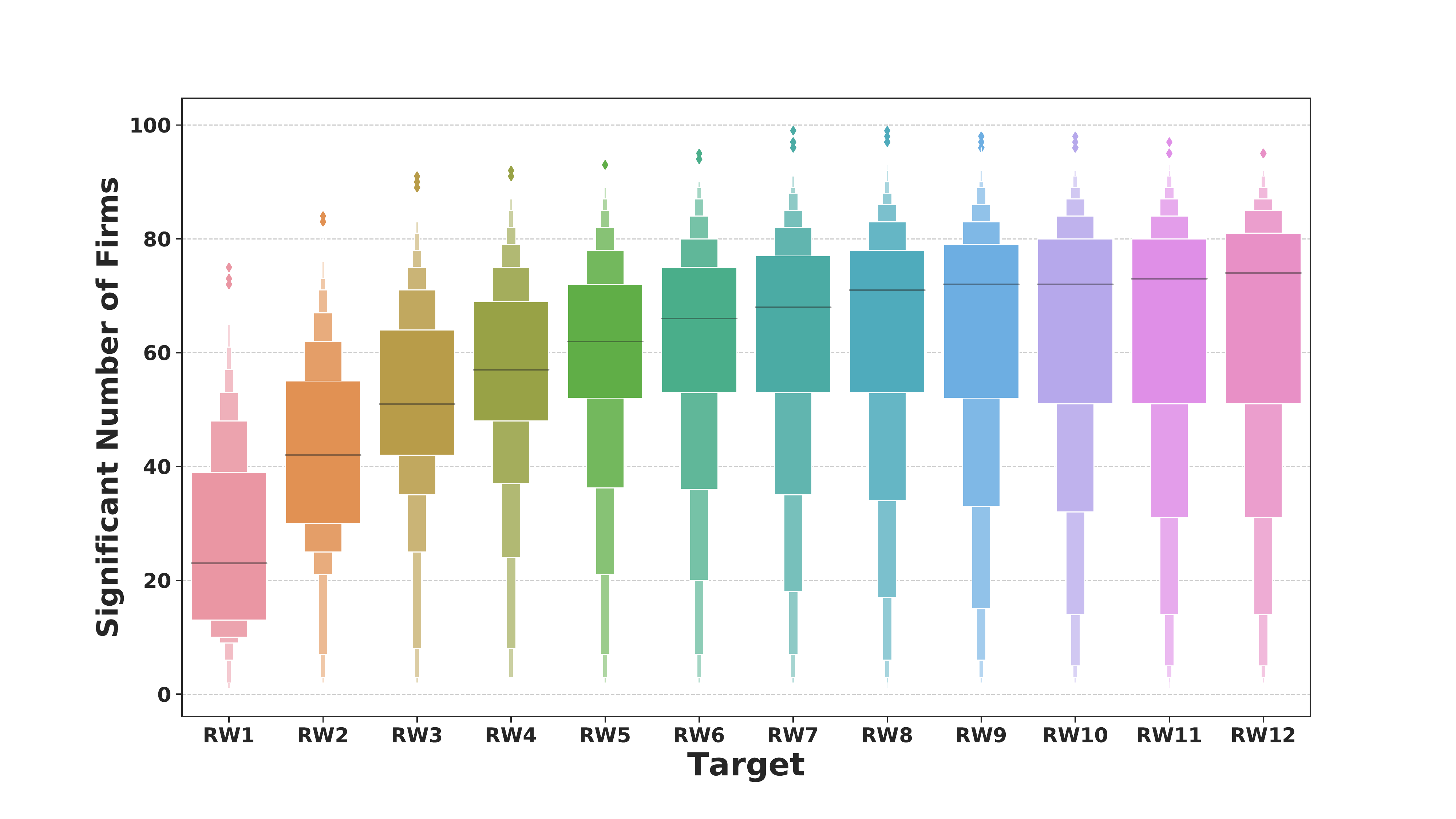} 
  \caption{{\bfseries The distribution of the significant number of firms for every feature with one specific target.}} 
  \label{target_siginicant_number_boxplot} 
\end{figure}

Aggregating by target, we draw a box chart of the significant number shown in Figure \ref{target_siginicant_number_boxplot}. In this figure, every box represents the significant numbers distribution of firms for all features with the specific target. For example, the largest number in $RW_8$ box is 99 of feature $W_9Client^{21}RatioDiffH_{12}$, which means that the relationship between target $RW_8$ and feature $W_9Client^{21}RatioDiffH_{12}$ in 99 out of 102 firms is significant. $RW_8$, representing eight weeks or almost two months of stock returns, was selected as our prediction target after we systematically compared all the quantiles of all targets, which is unexpectedly in line with the optimal lag length of word of mouth taking effect \citep{luo2009quantifying} and benefits from the delayed effect of customer reviews \citep{luo2009quantifying, tirunillai2012does, huang2018customer}.

\section{Models and Results}
In this section, based on the discretization definition method, regression problems of predicting the long-term stock returns are converted to corresponding classification problems. We execute feature selection to improve prediction performance. Both linear and nonlinear methods to solve the classification problems of stock returns are validated by 5-fold cross-validations on the training set for classifier selection, and we obtained a high-performance prediction model named XGB-OR. We then conduct a hold-out validation test with the selected model for realistic application evaluation; a baseline model based on ten technical indicators is also built for the further comparison.

\subsection{Preliminaries}
As illustrated in the previous sections, $RW_8$ is our target of prediction in this paper. In most cases, stockholders just care about the movement direction of stock price, which means that instead of the exact value of the long-term return, whether the return is positive or negative is the foremost interest in reality because it could provide advice on the direction of trading. Therefore, using zero as the cut-off point, $RW_{8,i}$ is transformed into a binary variable $y_i$ (the element on $i$-th week in $RW_8$), i.e., 
\begin{equation} 
y_i=\left\{
\begin{array}{ll}
1, &RW_{8,i}\geq 0 \\
0, &otherwise.
\end{array}
\right.
\label{return_discretization}
\end{equation}
Indeed, varying values of the cut-off point can test the sensitivity of prediction models, which will be examined later.

All features and targets are labeled with timestamps for the data alignment. Features are extracted from reviews before the timestamp, and target is the closing price change eight weeks after that timestamp. As shown in Figure \ref{comment_number}, the burst of online purchasing began around 2014; therefore, we divided the dataset into two parts according to the date: the training subset (from January 2014 to June 2017, 3.5 years) and the testing subset (from July 2017 to December 2017, half year). The training subset, which contains 14,688 firm-week samples, is used for feature selection among the extracted features and to fit and estimate the prediction model. The testing set, which contains 2,537 firm-week samples, is kept in a vault and brought out only at the end of evaluation for validating our model in realistic long-term (e.g., eight weeks) returns prediction.

\subsection{Feature Selection}
Irrelevant features can introduce considerable of noise, resulting in training the model towards a random wrong direction. Feature selection is one of the most important and frequently used techniques in data preprocessing for data mining \citep{blum1997selection}. It reduces the number of features, removes irrelevant, redundant, or noisy data, and brings immediate effects for applications: speeding up a data mining algorithm and improving mining performance such as predictive accuracy and result comprehensibility. Furthermore, in consideration of filter method being independent of any specific classifiers \citep{tang2014feature, blum1997selection}, after more than 7,000 features are extracted from the online review data, another remaining task is to select a subsample from them to achieve a competent performance in long-term returns prediction.

Currently, boosting \citep{freund1996experiments} is one of the best and therefore one of the most commonly employed classification methods in machine learning. This approach has been extensively discussed and analyzed, both in research and realistic application, and many different variants of boosting algorithms have been proposed for different purposes \citep{schapire1999improved, friedman2000additive}. Specifically, the boosting algorithm is capable of selecting the best combination of features \citep{creamer2010automated}. As one of the best boosting algorithms, GradientBoosting was chosen for the feature selection.

\begin{figure}[!htbp]
  \centering
  \includegraphics[width=\textwidth]{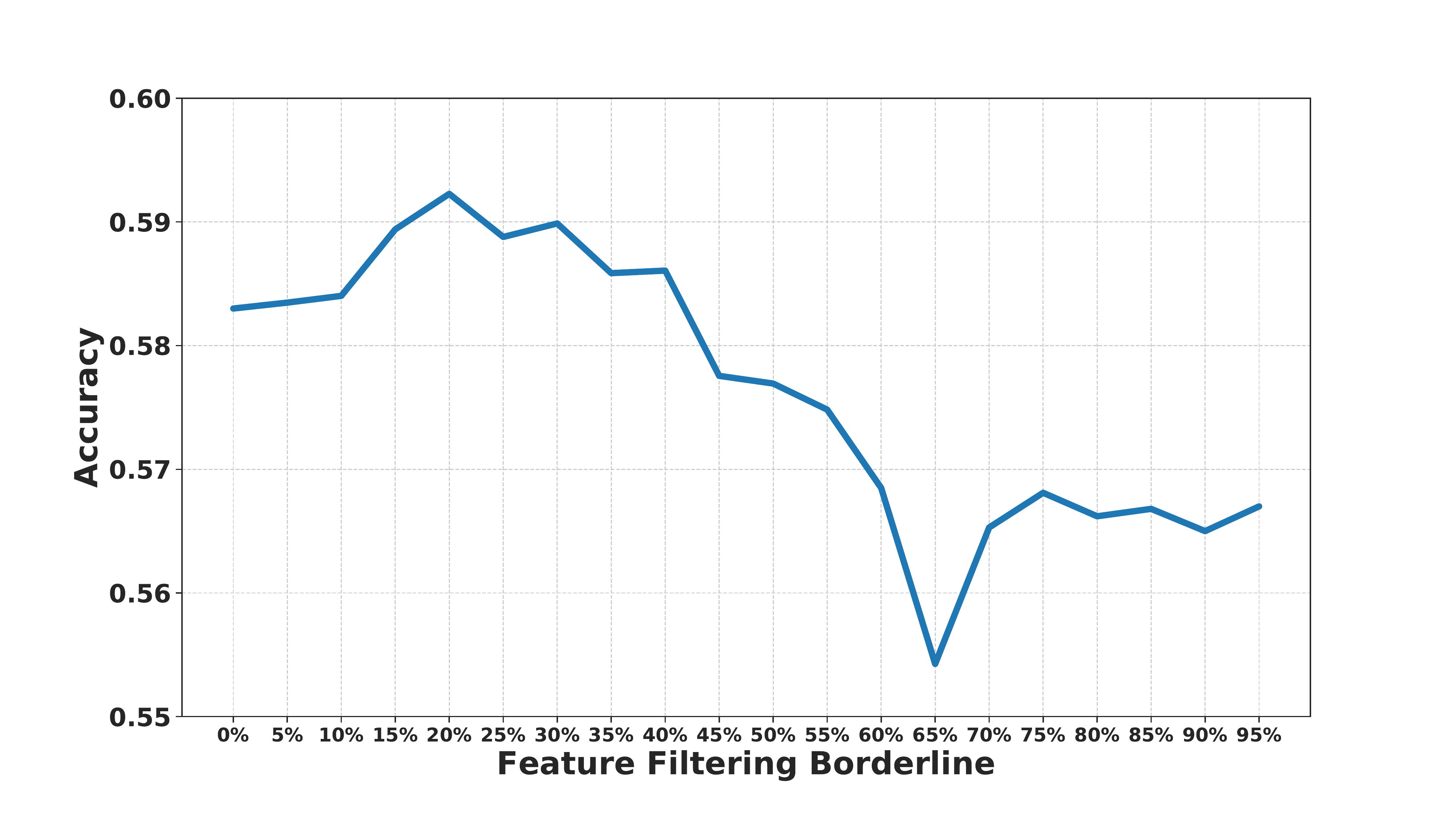} 
  \caption{{\bfseries Average accuracies of 5-fold cross-validation for different feature filtering borderlines.}} 
  \label{feature_selection} 
\end{figure}

In the above section, we count the significant number of firms for every feature-target pair. With the confirmed target as $RW_8$, through dividing the significant number by the number of sample firms, we obtain the passing rate for each feature $f$, which is denoted as $PassingRate_f=Count_{ft}/I$, where $t$ represents the selected target $RW_8$ and $I$ is the number of sample firms. We assume that valuable features are always significant with most firms' eight-week returns, that is, we can select features by restricting the $PassingRate$. For example, by setting the borderline to 0.2, we filter out features with $PassingRate$ less than 0.2. Then, through 5-fold cross-validation with the GradientBoosting classifier in the training subset, we obtain an average accuracy for those remaining features. By traversing $PassingRate$ from 0 to 0.95 with 0.05 step length, we conduct 5-fold cross-validations in feature selections. Finally, we obtain the average accuracies for each feature filtering shown in Figure \ref{feature_selection}. From this figure, it is clear that the average accuracies perform like a roller coaster; that is, the average accuracies rise and fall rapidly. The peak value of this trend line is 59.22\% with 0.2 borderline of $PassingRate$, and the accuracies finally decrease to the random value after peak. Accordingly, 6,246 features are selected for latter study after filtering out features with $PassingRate$ less than 0.2.

\subsection{Classifier Selection}
In this paper, we employ various machine learning algorithms, namely, XGB, GradientBoosting, AdaBoost, LSTM, Bagging, Logistic Regression, Random Forest and Gaussian Naive Bayes, to solve the classification problems for long-term stock return prediction. These methods are both cutting-edge and popular for training binary or multiple classification. To predict the categories (0, 1) of $y_i$ on $i$-th week, the input attributes of all models include all the features selected from the feature engineering. We adopt 5-fold cross-validation to systematically examine these models' performance in the training subset.

\begin{table}[!htbp]
\centering
\setlength{\belowcaptionskip}{10pt} 
\caption{Models Accuracy}
\label{tb_model_accuracy}
\resizebox{0.4\textwidth}{!}{
\begin{tabular}{ll}
\hline
Model & \multicolumn{1}{l}{Accuracy} \\ \hline
\textbf{XGB} & \textbf{59.65\%} \\
GradientBoosting & 59.22\% \\
AdaBoost & 57.75\% \\
LSTM & 57.05\% \\
\textbf{Random} & \textbf{56.77\%} \\
Bagging & 56.06\% \\
LogisticRegression & 55.73\% \\
RandomForest & 53.73\% \\
GaussianNB & 49.05\% \\ \hline
\end{tabular}
}
\end{table}

Accuracy, the most promising indicator in practical investments, is used to evaluate the performance of the proposed models. The accuracies of models by 5-fold cross-validation are shown in Table \ref{tb_model_accuracy}. Note that the \textbf{Random} means the percentage of the sample category with the largest proportion. Boosting algorithms and the deep learning model can beat the random line, especially the XGB model, which outperforms other models with the highest average accuracy. However, the remaining models do not even achieve an approving result. As a result, we choose the XGB model for our hold-out validation test and name it XGB-OR. Note that  to further test the effectiveness of feature selection, all classifiers were also tested with all features considered. We found that the best performance was significantly undermined by noisy features; in particular, the training process was much more time-consuming than the process with feature selection.

\subsection{Validation}
To further evaluate our prediction model in a more realistic manner, we apply our classification model for stock prediction on testing subset. Meanwhile, to avoid data snooping, we set the end date of training data to April 30, 2017. We evaluate the long-term stock return prediction with the selected XGB-OR model; the metrics are shown in Table \ref{tb_validation}. It turns out that the model achieves high prediction performance, with accuracy exceeding the random value (51.59\%) by almost 10\%. Moreover, the precision of 66.05\% in the positive direction compared with positive value (48.36\%) makes our model especially practical in realistic application. Except for XGB-OR, we also conducted a hold-out test with the other classifiers mentioned in the section above; there is no doubt that XGB-OR surpasses the other classifiers, presenting the highest accuracy.

\begin{table}[!htbp]
\centering
\setlength{\belowcaptionskip}{10pt} 
\caption{Validation}
\label{tb_validation}
\resizebox{0.4\textwidth}{!}{%
\begin{tabular}{lll}
\hline
 & XGB-OR & XGB-10TI \\ \hline
Accuracy & 61.02\% & 47.08\% \\
Precision & 66.05\% & 46.96\% \\
Recall & 37.73\% & 92.53\% \\
F-measure & 48.03\% & 62.30\% \\ \hline
\end{tabular}%
}
\end{table}

To further examine the robustness of predictive power of online product reviews, we compare the prediction performance of our model, which takes advantage of online reviews with the one taking technical indicators as features. Inspired by traditional financial time series forecasting approaches \citep{patel2015predicting, kim2003financial}, the baseline technical indicator model based on weekly prices, named XGB-10TI, is established. The input attributes of XGB-10TI are ten classical technical indicators shown in Table \ref{ten_indicator}. Furthermore, the prediction targets and the periods of training and testing set for two groups of models are identical. It can be seen from Table \ref{tb_validation} that the performance of the XGB-OR model is better than the baseline in both accuracy and precision. We can conclude that online reviews offer more predictive power than simple financial time series.

\subsection{Realistic application}
\begin{figure}[!htbp]
  \centering
  \includegraphics[width=1.\textwidth]{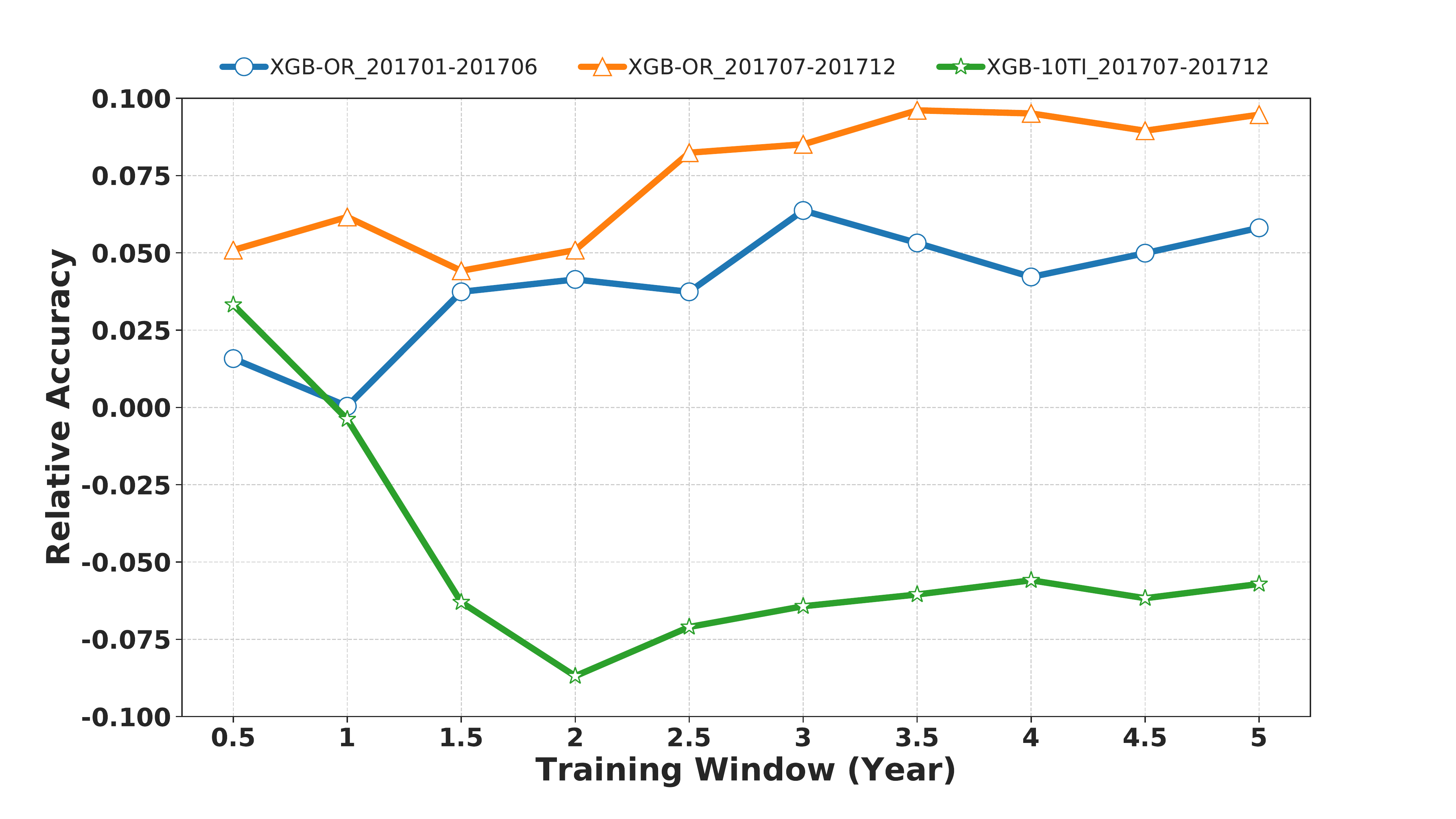} 
  \caption{{\bfseries Accuracies of hold-out tests with different training window lengths.} The relative accuracies represent the difference between predicting accuracies with different training windows and the random value of the predicting dataset.} 
  \label{holdout_window} 
\end{figure}

In the above section, we find that almost 3.5 years of online reviews have predictive power over the remaining half year of long-term stock returns. However, is the training data window perfect enough? For realistic application, we must determine the length of training data window to achieve a satisfactory result. We retain the testing dataset while expanding the training dataset from a half-year to five years, with a half-year being the step length used to locate the optimal training window size. For each training window, we conduct a hold-out test with our prediction model XGB-OR. To confirm the robustness, we shift the testing data window to an early period, which is from January 2017 to June 2017 with the same training windows. As shown in Figure \ref{holdout_window}, the relative accuracies indicate that more than three years of data are required to fully take advantage of online reviews.

In addition, we conduct training window selection on financial technical indicators. From Figure \ref{holdout_window}, the XGB-10TI\_201707-201712 line shows that only when the length of training data is a half year does the accuracy outperform the random value, which is consistent with the technical analysis in a short-term approach \citep{de1985does, jegadeesh1993returns, menkhoff2010use}. For the predicting period of January 2017 to June 2017, there is no training subset derived from sliding windows that can perform a higher accuracy than the random value; hence, we do not display the corresponding line in Figure \ref{holdout_window}.

\section{Discussion}
The competent performance of our approach, especially compared to solutions based on technique indicators, implies the unexpected power of online reviews in long-term prediction of stock returns. In particular, feeding more than three years of data to achieve the most satisfying accuracy suggests that online reviews in essence capture the feature regarding brand reputation and customer loyalty. In the consumer market, customers do not use the Explore and Exploit algorithm to go shopping but are used to buying products from familiar brands. Considering the long time required for brand building \cite{meenaghan1995role, urde2003core}, it takes time for consumers to build their attitude towards a new brand and consumption habit. Meanwhile, the impact of customer attitudes on firm equity is long-lasting, a finding supported by classical results from landmark studies \citep{aksoy2008long, mittal2005dual}. Specifically, it was already noted that the negative impact of online reviews can last 20 months, which is quite close to our ideal training data window \citep{luo2009quantifying}. From a theoretical point of view, in fact, our results also add to the existing literature: features regarding brand reputation or loyalty can be excellent indicators in long-term returns prediction, which has been missed in previous exploitations.

\begin{figure} 
  \centering 
  \subfigure[Sector ranking]{ 
    \includegraphics[width=.46\textwidth]{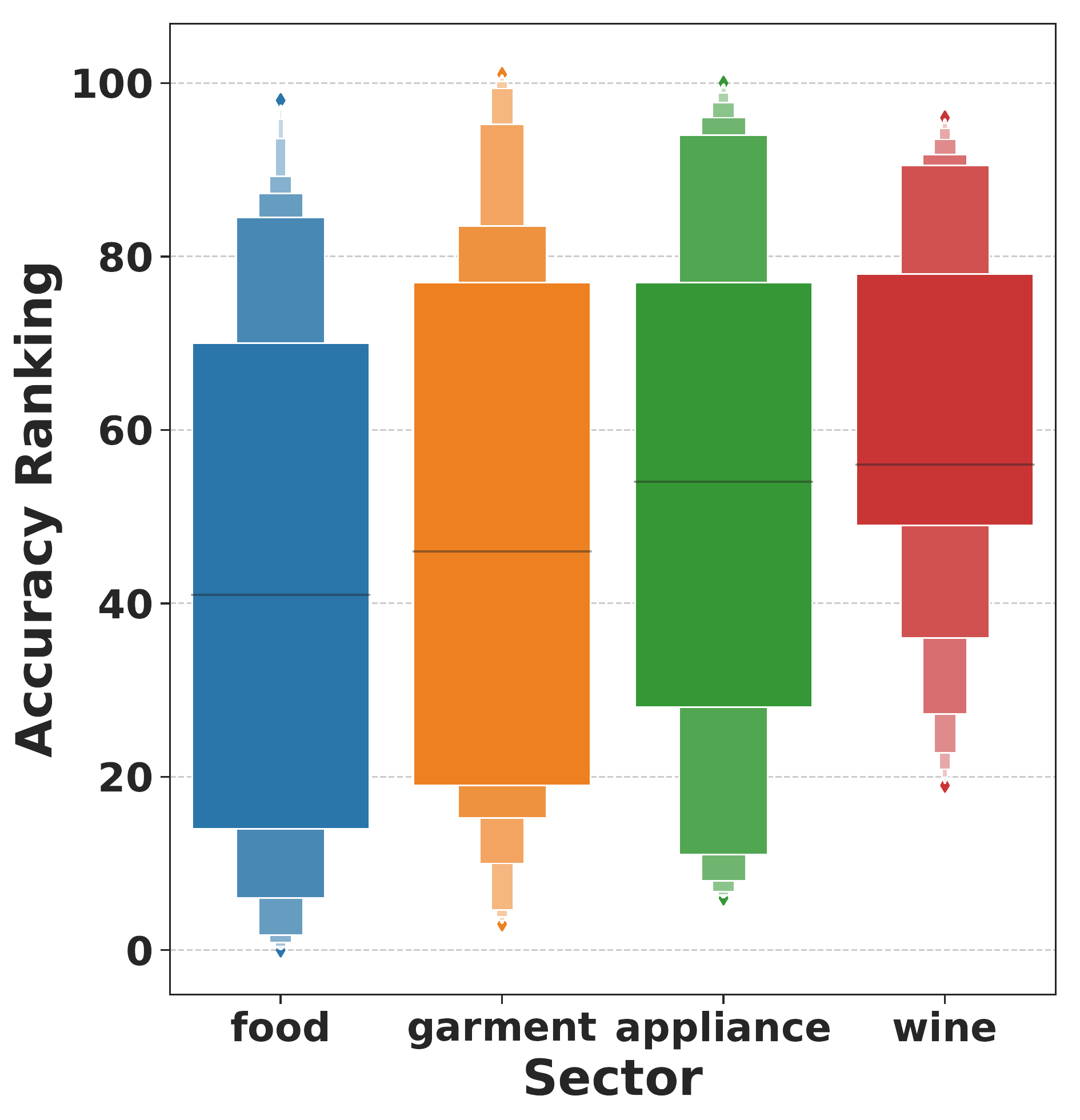} 
    \label{sector_ranking} 
  } 
  \subfigure[User proportion]{ 
    \includegraphics[width=.46\textwidth]{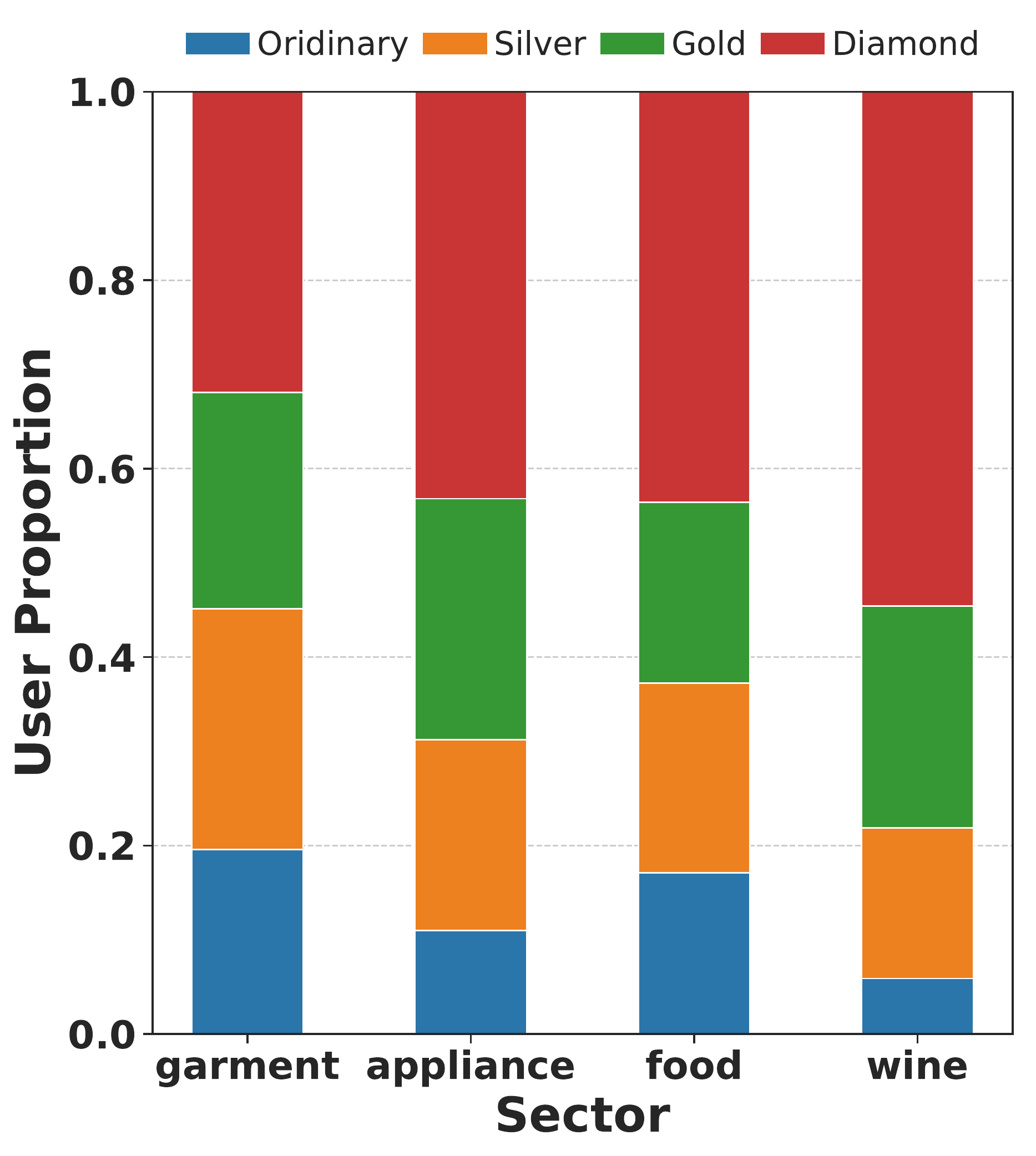} 
    \label{user_proportion} 
  } 
  \caption{{\bfseries Sector rankings and proportions of user-levels.}} 
\end{figure}

Prediction accuracy might vary across different stocks in our dataset because the impact from brand reputation or customer loyalty is intuitively diverse. As mentioned in our dataset, reviews of 102 companies were collected from four industry sectors. We calculate accuracies for all single firms and sort them. From the sorted list, we assign the rank position to each firm and draw the ranking box figure grouped by sector. As shown in Figure \ref{sector_ranking}, the wine sector outperforms all other sectors with the highest minimum and median rankings; this means that for all firms from the wine sector, our model achieves relatively high accuracy of long-term prediction. To determine what makes this difference, with the assumption that high level users generate more highly valuable information \citep{zhu2010impact}, we count the proportion of each type of user in every sector. There are four types of users, namely, ordinary users, silver users, gold users and diamond users. As shown in Figure \ref{user_proportion}, diamond users in the wine sector have the highest proportion, more than 50\%, and the percentage of diamond and gold users is approximately 80\%, which confirms our conjecture. These results remind us that prediction performance varying across individual stocks might undermine the previous efforts on only prediction of the market index, implying the necessity of establishing prediction solutions that are oriented to individual stocks. In particular, in the practical scenarios, most quantitative trading algorithms are based on individual stocks rather than indices, especially in countries that do not have sufficient financial tools for index trading.

\begin{figure}[!htbp]
  \centering
  \includegraphics[width=1.\textwidth]{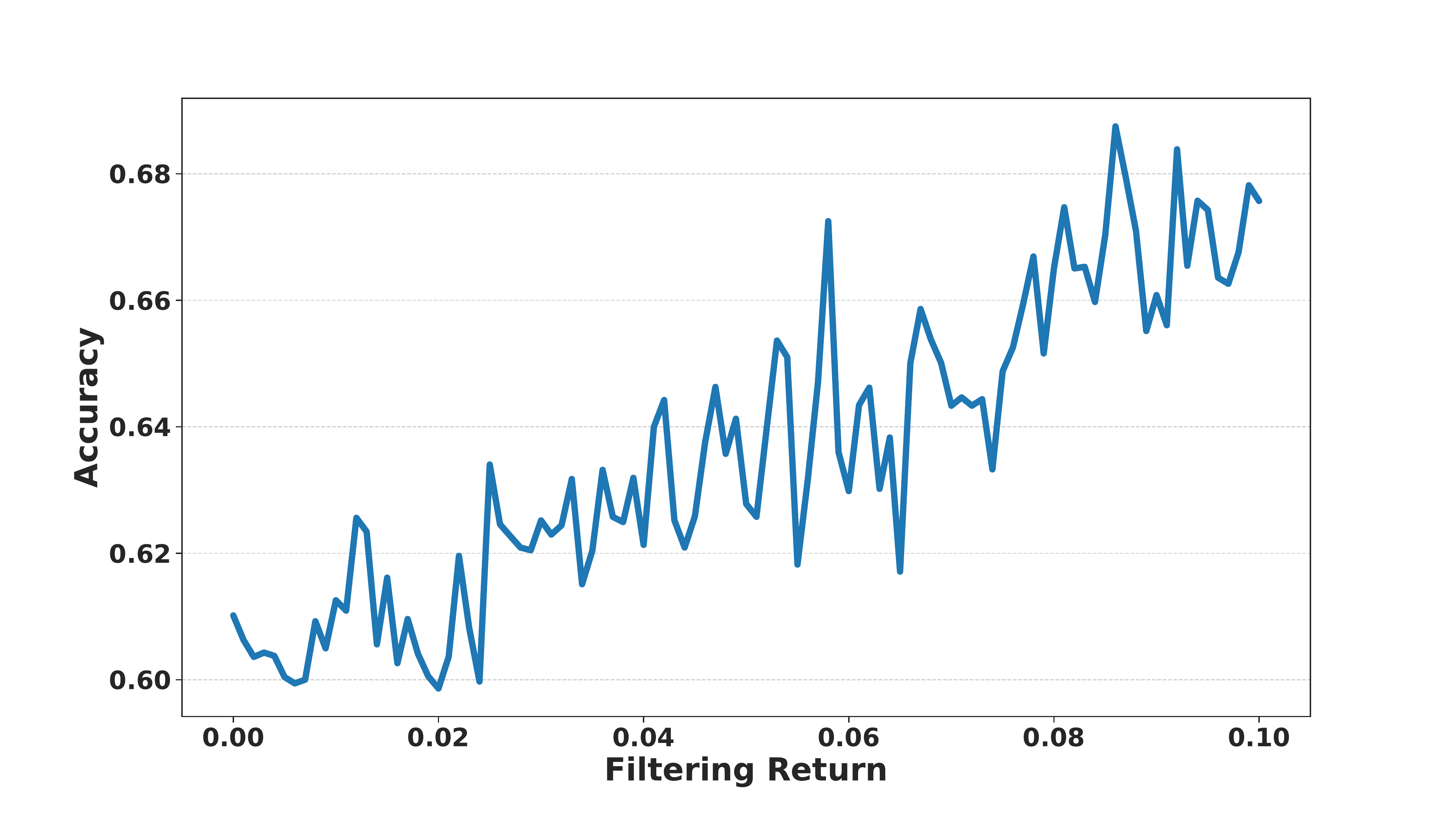} 
  \caption{{\bfseries Remove unstable samples to test the performance robustness.}} 
  \label{remove_unstable} 
\end{figure}

The cut-off point in determining the extent to which the return is positive or negative can reshape the establishment of the training and testing sets, accordingly influencing the evaluation of the prediction performance~\citep{wang2019aggregating}. To present a more comprehensive picture of performance valuation, extensions regarding more cut-off points are also further discussed. In the aforementioned model, we use zero as the cut-off point of the long-term stock return, which may be too sensitive to capture the significant difference between the dynamics of individual stocks. Therefore, to reduce the sensitivity, a simple threshold-based sampling approach presented in ~\citet{wang2019aggregating} is employed to further test the robustness of our approach. Define $\tau$ as the cut-off point of the eight-week close return, where $\tau \in [0, 0.1]$ with 0.001 step length \citep{wang2019aggregating}. For any given $\tau$, we filter out the observations with returns between -$\tau$ and $\tau$ and then conduct the hold-out test. For example, if $\tau = 0.03$, the samples with returns higher than 3\% or lower than -3\% are selected for predicting tests. Specifically, those samples with returns higher than 3\% are labeled, 1 and those with returns lower than -3\% are labeled 0. This approach has practical value in the real world, considering the transaction and impact cost, only above a certain level of return would be regarded as profitable. Moreover, the aim of displaying the threshold experiments is to further evaluate the prediction power and the robustness of our model. Figure \ref{remove_unstable} illustrates the relationship between $\tau$ and the accuracies. In this figure, with the growth of $\tau$, the accuracies display a general increasing tendency, which means that our model is more powerful for abnormal returns prediction. Our model reaches the minimum accuracy (59.86\%) with sample size $n = 14,506$ and $\tau = 0.02$ and the maximum accuracy (68.75\%) with sample size $n = 9,070$ and $\tau = 0.086$. However, the minimum accuracy still exceeds the random guess significantly, suggesting that the performance of our approach is robust and can be extensively used in different practical scenarios.

\section{Conclusion}
In this paper, we examine the prediction capability of online reviews. Using a large dataset of customer product reviews on JD.com, we find that long time review data positively predict long-term stock returns on an individual level. Meanwhile, the novel method of extracting features in our model differs from early studies and is proven to be effective in stock return prediction. The results in this paper highlight the role of consumers as information producers in financial markets. Compared with traditional information intermediaries such as technical indicators, consumer crowds can provide more persistent information on a company’s product and customer attitudes. Taken together, these findings provide evidence that the aggregated reviews of consumer crowds contain valuable information in long-term stock pricing.

There are inevitable limitations in the present study. For example, due to the limits of JD’s API, it is impossible to collect all product reviews, especially ones that were generated for historical products that are no longer being sold. Therefore, there might be information or survivorship bias in our model. Moreover, we did not put excessive effort into semantic analysis of texts in online reviews, which in fact might contain inspiring indicators in stock prediction. Both of the above limitations are promising directions in our future work.

\section*{Acknowledgments}
This work was supported by NSFC (Grant No. 71871006) and the fund of the State Key Lab of Software Development Environment (Grant No. SKLSDE-2017ZX-05).

\newpage
\appendix
\renewcommand{\appendixname}{Appendix}
\section{}

\setcounter{table}{0}
\setcounter{figure}{0}
\renewcommand{\thetable}{A.\arabic{table}}
\renewcommand{\thefigure}{A.\arabic{figure}}

\begin{figure}[!htbp]
  \centering
  \includegraphics[width=1.\textwidth]{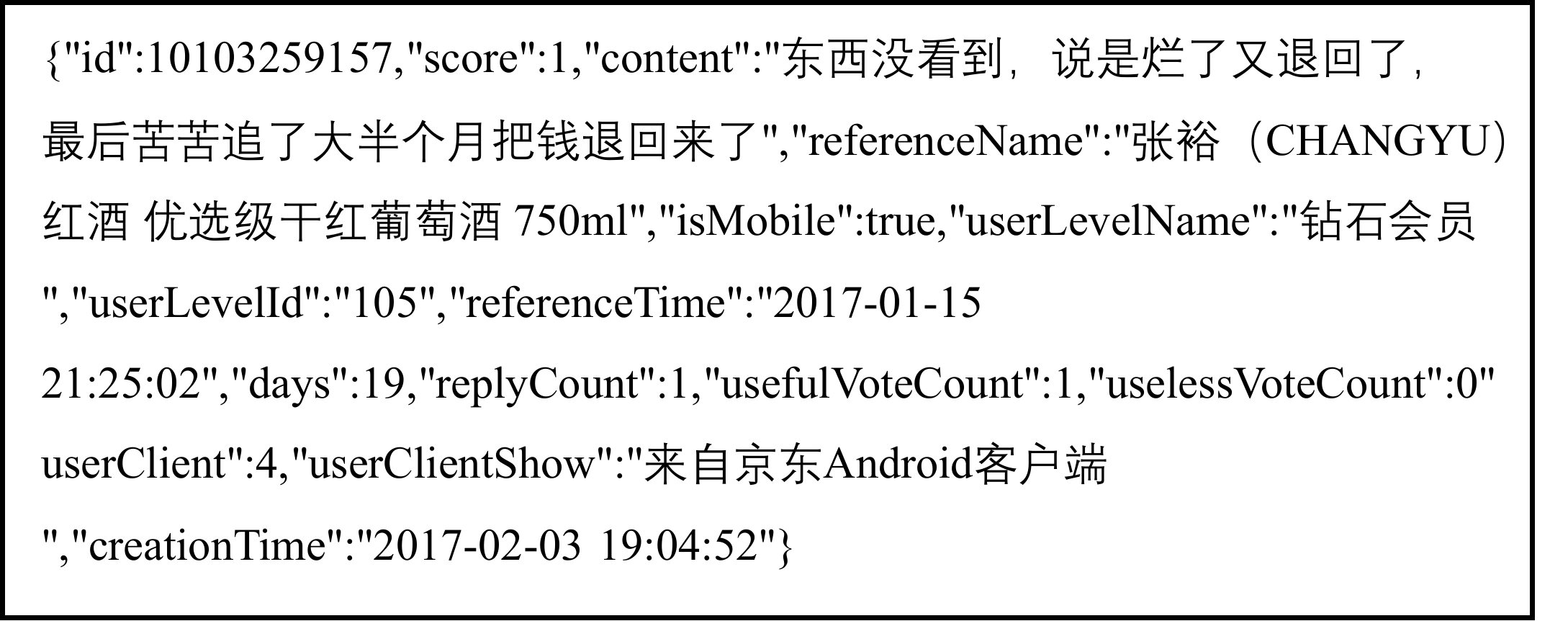}
  \caption{{\bfseries One online review sample from JD.com.}}
  \label{review_sample}
\end{figure}

\begin{figure}[!htbp]
  \centering
  \includegraphics[width=1.\textwidth]{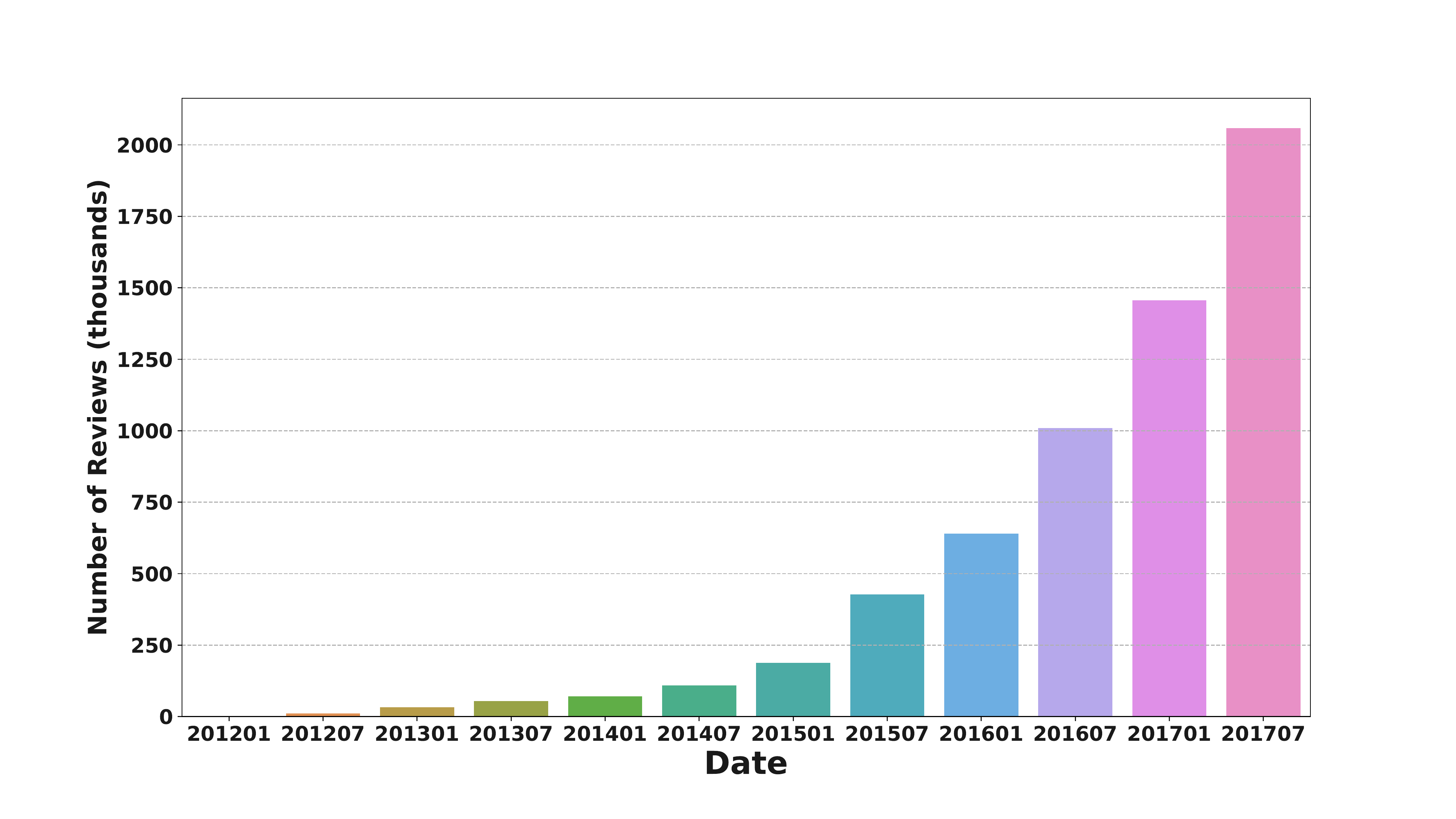}
  \caption{{\bfseries The number of all firms' product reviews in half-year granularity.} Around 2014, given the credit to ubiquity of mobile Internet industry, there is a exponential growth in online product reviews with the rapid rise of online shopping activities.}
  \label{comment_number}
\end{figure}

{\small
\begin{longtable}[c]{lllll}
\setlength{\belowcaptionskip}{10pt}\\
\caption{Review Firms}
\label{tb_firm_review}\\
\hline
Sectors                 & Codes    & Short Names          & \#Product & \#Review \\ \hline
\endfirsthead
\multicolumn{5}{c}%
{Table \thetable\ continued from previous page} \\
\hline
Sectors                 & Codes    & Short Names          & \#Product & \#Review \\ \hline
\endhead
\hline
\endfoot
\endlastfoot
Home Appliance & 002677SZ & ZHEJIANG MEIDA       & 74              & 2352           \\
Home Appliance & 600261SH & ZHEJIANG YANKON      & 165             & 2779           \\
Home Appliance & 600983SH & Whirlpool            & 248             & 2884           \\
Home Appliance & 002681SZ & FENDA                & 28              & 17035          \\
Home Appliance & 002076SZ & CNLIGHT              & 427             & 20118          \\
Home Appliance & 000921SZ & HISENSE HA           & 213             & 43347          \\
Home Appliance & 000533SZ & SHUNNA               & 930             & 48576          \\
Home Appliance & 002668SZ & HOMA                 & 69              & 55924          \\
Home Appliance & 002508SZ & ROBAM APPLIANCES     & 964             & 90841          \\
Home Appliance & 000418SZ & LITTLESWAN           & 1225            & 91222          \\
Home Appliance & 002543SZ & VANWARD ELECTRIC     & 473             & 93509          \\
Home Appliance & 600336SH & AUCMA                & 583             & 113729         \\
Home Appliance & 600060SH & HXDQ                 & 1525            & 128656         \\
Home Appliance & 000521SZ & CHML                 & 859             & 131582         \\
Home Appliance & 600839SH & CHANGHONG            & 1923            & 152903         \\
Home Appliance & 000541SZ & FSL                  & 649             & 154792         \\
Home Appliance & 000016SZ & KONKA GROUP          & 1784            & 204636         \\
Home Appliance & 002035SZ & VATTI                & 1387            & 246855         \\
Home Appliance & 000651SZ & GREE                 & 1593            & 310044         \\
Home Appliance & 002403SZ & ASD                  & 2297            & 405346         \\
Home Appliance & 600890SH & CRED HOLDING         & 5998            & 431188         \\
Home Appliance & 000100SZ & TCL                  & 5997            & 579173         \\
Home Appliance & 002242SZ & JOYOUNG              & 2677            & 673990         \\
Home Appliance & 000333SZ & MIDEA GROUP          & 5980            & 1580060        \\
Home Appliance & 002032SZ & SUPOR                & 5991            & 1664441        \\
Garment        & 002687SZ & GIUSEPPE             & 239             & 1395           \\
Garment        & 002569SZ & BUSEN CORP.          & 171             & 1513           \\
Garment        & 600107SH & mailyard             & 684             & 1663           \\
Garment        & 002574SZ & MING JEWELRY         & 453             & 2148           \\
Garment        & 002731SZ & CUIHUA JEWELRY       & 407             & 3239           \\
Garment        & 002044SZ & MEINIAN ONEHEALTH    & 596             & 5077           \\
Garment        & 002291SZ & SATURDAY             & 1480            & 5113           \\
Garment        & 600398SH & HLA                  & 5997            & 6811           \\
Garment        & 002036SZ & LIANCHUANG ELECTRON  & 36              & 7934           \\
Garment        & 600177SH & Youngor              & 3712            & 10602          \\
Garment        & 002612SZ & LANCY                & 1104            & 12465          \\
Garment        & 601718SH & Jihua Group          & 652             & 15264          \\
Garment        & 002003SZ & WEIXING              & 91              & 17575          \\
Garment        & 002269SZ & METERSBONWE          & 3213            & 25368          \\
Garment        & 002345SZ & CHJ                  & 1882            & 29763          \\
Garment        & 002397SZ & MENDALE              & 795             & 31116          \\
Garment        & 601566SH & JOEONE               & 4517            & 58651          \\
Garment        & 002563SZ & SEMIR                & 5060            & 74330          \\
Garment        & 600884SH & NBSS                 & 3794            & 76070          \\
Garment        & 603555SH & Guirenniao           & 6030            & 142494         \\
Garment        & 300005SZ & TOREAD               & 6000            & 207577         \\
Garment        & 002327SZ & FUANNA               & 1113            & 215161         \\
Garment        & 002293SZ & LUOLAI               & 5825            & 243486         \\
Garment        & 603001SH & AoKang               & 5956            & 244251         \\
Garment        & 600400SH & Hongdou Industrial   & 3974            & 293600         \\
Garment        & 600137SH & langshagufen         & 5996            & 1663330        \\
Garment        & 002029SZ & SEPTWOLVES           & 5997            & 1969384        \\
Food           & 002387SZ & VISIONOX             & 19              & 2832           \\
Food           & 000848SZ & CHENGDELOLO          & 17              & 4110           \\
Food           & 002481SZ & SHUANGTA FOOD        & 25              & 7390           \\
Food           & 002702SZ & HAIXIN FOODS         & 44              & 17560          \\
Food           & 002726SZ & LONGDA               & 25              & 21537          \\
Food           & 002330SZ & DELISI               & 138             & 22501          \\
Food           & 600419SH & TIAN RUN             & 335             & 27267          \\
Food           & 002515SZ & JINZI HAM            & 173             & 32537          \\
Food           & 600300SH & VVFB                 & 95              & 33546          \\
Food           & 002661SZ & KEMEN                & 101             & 40311          \\
Food           & 002507SZ & FULING ZHACAI        & 96              & 44595          \\
Food           & 002732SZ & YANTANG DAIRY        & 66              & 47143          \\
Food           & 600305SH & Hengshun Vinegar     & 184             & 47699          \\
Food           & 600597SH & BRIGHT DAIRY         & 238             & 48116          \\
Food           & 000639SZ & XIWANG               & 79              & 48205          \\
Food           & 002216SZ & SANQUAN FOODS        & 333             & 51833          \\
Food           & 000716SZ & NANFANG BLACK SESAME & 81              & 54021          \\
Food           & 600429SH & SANYUAN              & 496             & 70381          \\
Food           & 600073SH & SMAC                 & 430             & 76033          \\
Food           & 600298SH & ANGEL                & 322             & 105062         \\
Food           & 002695SZ & HUANGSHANGHUANG      & 400             & 117454         \\
Food           & 002570SZ & BEINGMATE            & 3369            & 143891         \\
Food           & 002582SZ & HAOXIANGNI           & 414             & 182331         \\
Food           & 000895SZ & SHUANGHUI            & 1085            & 185673         \\
Food           & 002557SZ & CHACHA FOOD CO.,LTD  & 402             & 191174         \\
Food           & 603288SH & HAI TIAN             & 159             & 194193         \\
Food           & 600887SH & YILI                 & 4454            & 488522         \\
Food           & 300146SZ & BY-HEALTH            & 3384            & 734398         \\
Food           & 600737SH & Cofco  Tunhe         & 6000            & 981302         \\
Wine           & 603369SH & King's Luck          & 348             & 2143           \\
Wine           & 600702SH & SHE DE               & 57              & 2779           \\
Wine           & 002461SZ & ZHUJIANG BREWERY     & 16              & 3476           \\
Wine           & 600199SH & AGSW                 & 68              & 5029           \\
Wine           & 600238SH & HAINAN YEDAO         & 51              & 11538          \\
Wine           & 600543SH & GSMG                 & 169             & 16934          \\
Wine           & 600197SH & YLT                  & 173             & 17063          \\
Wine           & 601579SH & KUAIJI               & 729             & 18894          \\
Wine           & 600779SH & SCSF                 & 396             & 24856          \\
Wine           & 000729SZ & YANJING BREWERY      & 103             & 25121          \\
Wine           & 000799SZ & JGJC                 & 682             & 27919          \\
Wine           & 600559SH & lao bai gan jiu      & 311             & 28326          \\
Wine           & 600600SH & TSINGTAO BREWERY     & 139             & 40665          \\
Wine           & 000596SZ & GUJING DISTILLERY    & 570             & 50731          \\
Wine           & 600059SH & GYLS                 & 1210            & 53641          \\
Wine           & 000869SZ & CHANGYU              & 723             & 80161          \\
Wine           & 600809SH & Shanxi Fen Wine      & 1786            & 136564         \\
Wine           & 600519SH & KWEICHOW MOUTAI      & 5555            & 137011         \\
Wine           & 002304SZ & YANGHE               & 1686            & 158968         \\
Wine           & 000568SZ & LUZHOU LAO JIAO      & 3131            & 245394         \\
Wine           & 000858SZ & WULIANGYE            & 4015            & 288223         \\ \hline
\end{longtable}
}

\begin{table}[!htbp]
\centering
\setlength{\belowcaptionskip}{10pt}
\caption{Variant Features}
\label{tb_features_variant}
\resizebox{\textwidth}{!}{
\begin{tabular}{ll} \hline
Feature & Definition \\ \hline
$W_nBasicFeatureDiff$ & $ W_nBasicFeature_i - W_nBasicFeature_{i-1} $ \\
$W_nBasicFeatureDiffRatio$ & $ \frac{W_nBasicFeature_{i} - W_nBasicFeature_{i-1}}{W_nBasicFeature_{i-1}} $ \\
$W_nBasicFeatureRatio$ & $ \frac{W_nBasicFeature_i}{W_nReview_i} $ \\
$W_nBasicFeatureRatioDiff$ & $ W_nBasicFeatureRatio_{i} - W_nBasicFeatureRatio_{i-1} $ \\
$W_nBasicFeatureDiffH_m$ & $ W_nBasicFeature_{i} - W_mBasicFeature_{i-n} $ \\
$W_nBasicFeatureRatioDiffH_m$ & $ W_nBasicFeatureRatio_{i} - W_mBasicFeatureRatio_{i-n} $ \\ 
$W_nBasicFeatureAverage$ & $ \frac{W_nBasicFeature_i}{W_nReview_i} $ \\
$W_nBasicFeatureAverageDiff$ & $ W_nBasicFeatureAverage_{i} - W_nBasicFeatureAverage_{i-1} $ \\
$W_nBasicFeatureAverageDiffH_m$ & $ W_nBasicFeatureAverage_{i} - W_mBasicFeatureAverage_{i-n} $ \\ \hline
\end{tabular}}
\footnotesize{
\begin{flushleft}
The $Diff$ variant is the difference sequence of each feature series. The $DiffRatio$ variant represents the rate of change between two adjacent time points. The $Ratio$ variant is the proportion of reviews with certain features in $n$ weeks. The $RatioDiff$ variant represents the proportion change between two adjacent time points. The $DiffH_m$ variant is the difference between values of certain features in $n$ weeks and that in the prior $m$ weeks. The $RatioDiffH_m$ variant represents the difference between the proportion of reviews with some feature in $n$ weeks and that in the prior $m$ weeks. The formula of $Average$ variant is the same with $Ratio$ variant, but indicates the average value of reviews in some feature within $n$ weeks, such as $W_nUsefulAverage$ means the average number of useful vote for every review in $n$ weeks. The $AverageDiff$ variant shows the average value change between two adjacent time points. The $AverageDiffH_m$ variant expresses the difference between the average value of reviews with some feature in $n$ weeks and that in the prior $m$ weeks.
\end{flushleft}}
\end{table}

\begin{table}[!htbp]
\centering
\setlength{\belowcaptionskip}{10pt}
\caption{Review Features}
\label{tb_features_review}
\resizebox{0.6\textwidth}{!}{
\begin{tabular}{ll} \hline
Feature & Definition \\ \hline
$W_nReview$ & $ N $ \\ 
$W_nReviewDiff$ & $ W_nReview_i - W_nReview_{i-1} $ \\
$W_nReviewDiffRatio$ & $ \frac{W_nReview_{i} - W_nReview_{i-1}}{W_nReview_{i-1}} $ \\
$W_nReviewDiffH_m$ & $ W_nReview_{i} - W_mReview_{i-n} $ \\\hline
\end{tabular}}
\end{table}

\begin{table}[!htbp]
\centering
\setlength{\belowcaptionskip}{10pt}
\caption{Star Features}
\label{tb_features_star}
\resizebox{0.8\textwidth}{!}{
\begin{tabular}{ll} \hline
Feature & Definition \\ \hline
$W_nStar^s$ & $ \sum_{i=1}^N\{1\ if\ star_i==s\ else\ 0\} $ \\ 
$W_nStar^sDiff$ & $ W_nStar^s_i - W_nStar^s_{i-1} $ \\
$W_nStar^sDiffRatio$ & $ \frac{W_nStar^s_{i} - W_nStar^s_{i-1}}{W_nStar^s_{i-1}} $ \\
$W_nStar^sRatio$ & $ \frac{W_nStar^s_i}{W_nReview_i} $ \\
$W_nStar^sRatioDiff$ & $ W_nStar^sRatio_{i} - W_nStar^sRatio_{i-1} $ \\
$W_nStar^sDiffH_m$ & $ W_nStar^s_{i} - W_mStar^s_{i-n} $ \\
$W_nStar^sRatioDiffH_m$ & $ W_nStar^sRatio_{i} - W_mStar^sRatio_{i-n} $ \\
$W_nStar^{15}Diff$ & $ W_nStar^5_i - W_nStar^1_i $ \\
$W_nStar^{15}DiffRatio$ & $ \frac{W_nStar^{15}Diff_i-W_nStar^{15}Diff_{i-1}}{W_nStar^{15}Diff_{i-1}} $ \\
$W_nStar^{15}Ratio$ & $ W_nStar^5Ratio_i - W_nStar^1Ratio_i $ \\
$W_nStar^{15}RatioDiff$ & $ W_nStar^{15}Ratio_{i} - W_nStar^{15}Ratio_{i-1} $ \\
\hline
\end{tabular}}
\end{table}

\begin{table}[!htbp]
\centering
\setlength{\belowcaptionskip}{10pt}
\caption{Default Features}
\label{tb_features_default}
\resizebox{\textwidth}{!}{
\begin{tabular}{ll}
\hline
Feature & Definition \\ \hline
$W_nDefault$ & $ \sum_{i=1}^N\{1\ if\ isDefault_i==True\ else\ 0 \}$ \\
$W_nDefaultDiff$ & $ W_nDefault_i - W_nDefault_{i-1} $ \\
$W_nDefaultDiffRatio$ & $ \frac{W_nDefault_{i} - W_nDefault_{i-1}}{W_nDefault_{i-1}} $ \\
$W_nDefaultRatio$ & $ \frac{W_nDefault_i}{W_nReview_i} $ \\
$W_nDefaultRatioDiff$ & $ W_nDefaultRatio_{i} - W_nDefaultRatio_{i-1} $ \\
$W_nDefaultDiffH_m$ & $ W_nDefault_{i} - W_mDefault_{i-n} $ \\
$W_nDefaultRatioDiffH_m$ & $ W_nDefaultRatio_{i} - W_mDefaultRatio_{i-n} $ \\ 
\hline
\end{tabular}}
\end{table}

\begin{table}[!htbp]
\centering
\setlength{\belowcaptionskip}{10pt}
\caption{Score Features}
\label{tb_features_score}
\resizebox{0.6\textwidth}{!}{
\begin{tabular}{ll} \hline
Feature & Definition \\ \hline
$W_nScore$ & $ \frac{\sum_{s=1}^5(s\times W_nStar^s_i)}{W_nReview_i} $ \\ 
$W_nScoreDiff$ & $ W_nScore_i - W_nScore_{i-1} $ \\
$W_nScoreDiffRatio$ & $ \frac{W_nScore_{i} - W_nScore_{i-1}}{W_nScore_{i-1}} $ \\
$W_nScoreDiffH_m$ & $ W_nScore_{i} - W_mScore_{i-n} $\\\hline
\end{tabular}}
\end{table}

\begin{table}[!htbp]
\centering
\setlength{\belowcaptionskip}{10pt}
\caption{Emotion Variant Features}
\label{tb_features_emotionVariant}
\resizebox{\textwidth}{!}{
\begin{tabular}{ll} \hline
Feature & Definition \\ \hline
$W_nEmotionFeatureRatioE$ & $ \frac{W_nEmotionFeature}{W_nEmotion} $ \\
$W_nEmotionFeatureRatioEDiff$ & $ W_nEmotionFeatureRatioE_i - W_nEmotionFeatureRatioE_{i-1} $ \\
$W_nEmotionFeatureRatioEDiffH_m$ & $ W_nEmotionFeatureRatioE_{i} - W_mEmotionFeatureRatioE_{i-n} $\\\hline
\end{tabular}}
\footnotesize{
\begin{flushleft}
The $RatioE$ variant represents the proportion of reviews with certain emotions within emotional reviews in $n$ weeks and the other two have the same meaning with $RatioDiff$ and $RatioDiffH_m$ however the number of emotional reviews $W_nEmotion$ as denominator. These three special variants are only applicable to $W_nEmotion^e$ and $W_nEmotion^{negative}$.
\end{flushleft}}
\end{table}

\begin{table}[!htbp]
\centering
\setlength{\belowcaptionskip}{10pt}
\caption{Emotion Features}
\label{tb_features_emotion}
\resizebox{\textwidth}{!}{
\begin{tabular}{ll} \hline
Feature & Definition \\ \hline
$W_nEmotion^e$ & $ \sum_{i=1}^N\{1\ if\ ReviewEmotion_i==e\ else\ 0\} $ \\ 
$W_nEmotion^eDiff$ & $ W_nEmotion^e_i - W_nEmotion^e_{i-1} $ \\
$W_nEmotion^eDiffRatio$ & $ \frac{W_nEmotion^e_{i} - W_nEmotion^e_{i-1}}{W_nEmotion^e_{i-1}} $ \\
$W_nEmotion^eRatio$ & $ \frac{W_nEmotion^e}{W_nReview} $ \\
$W_nEmotion^eRatioDiff$ & $ W_nEmotion^eRatio_i - W_nEmotion^eRatio_{i-1} $ \\
$W_nEmotion^eDiffH_m$ & $ W_nEmotion^e_{i} - W_mEmotion^e_{i-n} $\\
$W_nEmotion^eRatioDiffH_m$ & $ W_nEmotion^eRatio_{i} - W_mEmotion^eRatio_{i-n} $\\

$W_nEmotion$ & $ \sum_{e=0}^4{W_nEmotion^e} $ \\
$W_nEmotion^eRatioE$ & $ \frac{W_nEmotion^e}{W_nEmotion} $ \\
$W_nEmotion^eRatioEDiff$ & $ W_nEmotion^eRatioE_i - W_nEmotion^eRatioE_{i-1} $ \\
$W_nEmotion^eRatioEDiffH_m$ & $ W_nEmotion^eRatioE_{i} - W_mEmotion^eRatioE_{i-n} $\\

$W_nEmotionDiff$ & $ W_nEmotion_i - W_nEmotion_{i-1} $ \\
$W_nEmotionDiffRatio$ & $ \frac{W_nEmotion_{i} - W_nEmotion_{i-1}}{W_nEmotion_{i-1}} $ \\
$W_nEmotionRatio$ & $ \frac{W_nEmotion}{W_nReview} $ \\
$W_nEmotionRatioDiff$ & $ W_nEmotionRatio_i - W_nEmotionRatio_{i-1} $ \\
$W_nEmotionDiffH_m$ & $ W_nEmotion_{i} - W_mEmotion_{i-n} $\\
$W_nEmotionRatioDiffH_m$ & $ W_nEmotionRatio_{i} - W_mEmotionRatio_{i-n} $\\

$W_nEmotion^{negative}$ & $ \sum_{i=1}^N\{1\ if\ ReviewEmotion_i\in\{0, 1, 3, 4\}\ else\ 0\} $ \\
$W_nEmotion^{negative}Diff$ & $ W_nEmotion^{negative}_i - W_nEmotion^{negative}_{i-1} $ \\
$W_nEmotion^{negative}DiffRatio$ & $ \frac{W_nEmotion^{negative}_{i} - W_nEmotion^{negative}_{i-1}}{W_nEmotion^{negative}_{i-1}} $ \\
$W_nEmotion^{negative}Ratio$ & $ \frac{W_nEmotion^{negative}}{W_nReview} $ \\
$W_nEmotion^{negative}RatioDiff$ & $ W_nEmotion^{negative}Ratio_i - W_nEmotion^{negative}Ratio_{i-1} $ \\
$W_nEmotion^{negative}DiffH_m$ & $ W_nEmotion^{negative}_{i} - W_mEmotion^{negative}_{i-n} $\\
$W_nEmotion^{negative}RatioDiffH_m$ & $ W_nEmotion^{negative}Ratio_{i} - W_mEmotion^{negative}Ratio_{i-n} $\\
$W_nEmotion^{negative}RatioE$ & $ \frac{W_nEmotion^{negative}}{W_nEmotion} $ \\
$W_nEmotion^{negative}RatioEDiff$ & $ W_nEmotion^{negative}RatioE_i - W_nEmotion^{negative}RatioE_{i-1} $ \\
$W_nEmotion^{negative}RatioEDiffH_m$ & $ W_nEmotion^{negative}RatioE_{i} - W_mEmotion^{negative}RatioE_{i-n} $\\
\hline
\end{tabular}}
\end{table}

{\small
\begin{longtable}{ll}
\setlength{\belowcaptionskip}{10pt} \\
\caption{Tendency Features}
\label{tb_features_tendency} \\
\hline
Feature & Definition \\ \hline
\endfirsthead
\multicolumn{2}{c}%
{ Table \thetable\ continued from previous page} \\
\hline
Feature & Definition \\ \hline
\endhead
\hline
\endfoot
\endlastfoot
$W_nTendency^{posW}$ & $ \sum_{i=1}^NReviewPos_i $ \\
$W_nTendency^{posW}Diff$ & $ W_nTendency^{posW}_i - W_nTendency^{posW}_{i-1} $ \\
$W_nTendency^{posW}DiffRatio$ & $ \frac{W_nTendency^{posW}_{i} - W_nTendency^{posW}_{i-1}}{W_nTendency^{posW}_{i-1}} $ \\
$W_nTendency^{posW}Average$ & $ \frac{W_nTendency^{posW}}{W_nReview} $ \\
$W_nTendency^{posW}AverageDiff$ & $ W_nTendency^{posW}Average_i - W_nTendency^{posW}Average_{i-1} $ \\
$W_nTendency^{posW}DiffH_m$ & $ W_nTendency^{posW}_{i} - W_mTendency^{posW}_{i-n} $ \\
$W_nTendency^{posW}AverageDiffH_m$ & $ W_nTendency^{posW}Average_{i} - W_mTendency^{posW}Average_{i-n} $ \\
$W_nTendency^{negW}$ & $ \sum_{i=0}^NReviewNeg_i $ \\
$W_nTendency^{negW}Diff$ & $ W_nTendency^{negW}_i - W_nTendency^{negW}_{i-1} $ \\
$W_nTendency^{negW}DiffRatio$ & $ \frac{W_nTendency^{negW}_{i} - W_nTendency^{negW}_{i-1}}{W_nTendency^{negW}_{i-1}} $ \\
$W_nTendency^{negW}Average$ & $ \frac{W_nTendency^{negW}}{W_nReview} $ \\
$W_nTendency^{negW}AverageDiff$ & $ W_nTendency^{negW}Ratio_i - W_nTendency^{negW}Ratio_{i-1} $ \\
$W_nTendency^{negW}DiffH_m$ & $ W_nTendency^{negW}_{i} - W_mTendency^{negW}_{i-n} $ \\
$W_nTendency^{negW}AverageDiffH_m$ & $ W_nTendency^{negW}Average_{i} - W_mTendency^{negW}Average_{i-n} $ \\
$W_nTendency^{word}$ & $ \frac{W_nTendency_{pos}-W_nTendency^{neg}}{W_nTendency_{pos}+W_nTendency^{neg}} $ \\
$W_nTendency^{word}Diff$ & $ W_nTendency^{word}_i - W_nTendency^{word}_{i-1} $ \\
$W_nTendency^{word}DiffH_m$ & $ W_nTendency^{word}_i - W_nTendency^{word}_{i-n} $ \\
$W_nTendency^{posR}$ & $ \sum_{i=0}^N\{1\ if\ ReviewPos_i>ReviewNeg_i\ else\ 0\} $ \\
$W_nTendency^{posR}Diff$ & $ W_nTendency^{posR}_i - W_nTendency^{posR}_{i-1} $ \\
$W_nTendency^{posR}DiffRatio$ & $ \frac{W_nTendency^{posR}_{i} - W_nTendency^{posR}_{i-1}}{W_nTendency^{posR}_{i-1}} $ \\
$W_nTendency^{posR}Ratio$ & $ \frac{W_nTendency^{posR}}{W_nReview} $ \\
$W_nTendency^{posR}RatioDiff$ & $ W_nTendency^{posR}Ratio_i - W_nTendency^{posR}Ratio_{i-1} $ \\
$W_nTendency^{posR}DiffH_m$ & $ W_nTendency^{posR}_{i} - W_mTendency^{posR}_{i-n} $ \\
$W_nTendency^{posR}RatioDiffH_m$ & $ W_nTendency^{posR}Ratio_{i} - W_mTendency^{posR}Ratio_{i-n} $ \\
$W_nTendency^{negR}$ & $ \sum_{i=0}^N\{1\ if\ ReviewPos_i<ReviewNeg_i\ else\ 0\} $ \\
$W_nTendency^{negR}Diff$ & $ W_nTendency^{negR}_i - W_nTendency^{negR}_{i-1} $ \\
$W_nTendency^{negR}DiffRatio$ & $ \frac{W_nTendency^{negR}_{i} - W_nTendency^{negR}_{i-1}}{W_nTendency^{negR}_{i-1}} $ \\
$W_nTendency^{negR}Ratio$ & $ \frac{W_nTendency^{negR}}{W_nReview} $ \\
$W_nTendency^{negR}RatioDiff$ & $ W_nTendency^{negR}Ratio_i - W_nTendency^{negR}Ratio_{i-1} $ \\
$W_nTendency^{negR}DiffH_m$ & $ W_nTendency^{negR}_{i} - W_mTendency^{negR}_{i-n} $ \\
$W_nTendency^{negR}RatioDiffH_m$ & $ W_nTendency^{negR}Ratio_{i} - W_mTendency^{negR}Ratio_{i-n} $ \\
$W_nTendency^{pos}$ & $ \sum_{i=0}^N\{ReviewTen_i \ if\ ReviewPos_i>ReviewNeg_i\ else\ 0\} $ \\
$W_nTendency^{pos}Diff$ & $ W_nTendency^{pos}_i - W_nTendency^{pos}_{i-1} $ \\
$W_nTendency^{pos}DiffRatio$ & $ \frac{W_nTendency^{pos}_{i} - W_nTendency^{pos}_{i-1}}{W_nTendency^{pos}_{i-1}} $ \\
$W_nTendency^{pos}Average$ & $ \frac{W_nTendency^{pos}}{W_nTendency^{posR}} $ \\
$W_nTendency^{pos}AverageDiff$ & $ W_nTendency^{pos}Average_i - W_nTendency^{pos}Average_{i-1} $ \\
$W_nTendency^{pos}DiffH_m$ & $ W_nTendency^{pos}_{i} - W_mTendency^{pos}_{i-n} $ \\
$W_nTendency^{pos}AverageDiffH_m$ & $ W_nTendency^{pos}Average_{i} - W_mTendency^{pos}Average_{i-n} $ \\
$W_nTendency^{neg}$ & $ \sum_{i=0}^N\{ReviewTen_i \ if\ ReviewPos_i<ReviewNeg_i\ else\ 0\} $ \\
$W_nTendency^{neg}Diff$ & $ W_nTendency^{neg}_i - W_nTendency^{neg}_{i-1} $ \\
$W_nTendency^{neg}DiffRatio$ & $ \frac{W_nTendency^{neg}_{i} - W_nTendency^{neg}_{i-1}}{W_nTendency^{neg}_{i-1}} $ \\
$W_nTendency^{neg}Average$ & $ \frac{W_nTendency^{neg}}{W_nTendency^{negR}} $ \\
$W_nTendency^{neg}AverageDiff$ & $ W_nTendency^{neg}Average_i - W_nTendency^{neg}Average_{i-1} $ \\
$W_nTendency^{neg}DiffH_m$ & $ W_nTendency^{neg}_{i} - W_mTendency^{neg}_{i-n} $ \\
$W_nTendency^{neg}AverageDiffH_m$ & $ W_nTendency^{neg}Average_{i} - W_mTendency^{neg}Average_{i-n} $ \\
$W_nTendency$ & $ W_nTendency^{pos} + W_nTendency^{neg} $ \\
$W_nTendencyDiff$ & $ W_nTendency_i - W_nTendency_{i-1} $ \\
$W_nTendencyDiffRatio$ & $ \frac{W_nTendency_{i} - W_nTendency_{i-1}}{W_nTendency_{i-1}} $ \\
$W_nTendencyDiffH_m$ & $ W_nTendency_{i} - W_mTendency_{i-n} $ \\ \hline
\end{longtable}
}

\begin{table}[!htbp]
\centering
\setlength{\belowcaptionskip}{10pt}
\caption{Days Features}
\label{tb_features_days}
\resizebox{0.8\textwidth}{!}{
\begin{tabular}{ll}
\hline
Feature & Definition \\ \hline
$W_nDays$ & $ \frac{\sum_{i=0}^Ndays_i}{W_nReview}$ \\
$W_nDaysDiff$ & $ W_nDays_{i} - W_nDays_{i-1}$ \\
$W_nDaysDiffRatio$ & $ \frac{W_nDays_{i} - W_nDays_{i-1}}{W_nDays_{i-1}}$ \\
$W_nDaysH_m$ & $ W_nDays_{i}\times W_nReview_i - W_mDays_{i-n}\times W_mReview_{i-n}$ \\
$W_nDaysDiffH_m$ & $ W_nDays_{i} - W_mDays_{i-n} $ \\ \hline
\end{tabular}}
\end{table}

\begin{table}[!htbp]
\centering
\setlength{\belowcaptionskip}{10pt}
\caption{Useful Features}
\label{tb_features_useful}
\resizebox{\textwidth}{!}{
\begin{tabular}{ll}
\hline
Feature & Definition \\ \hline
$W_nUseful$ & $ \sum_{i=1}^NusefulVoteCount_i$ \\
$W_nUsefulDiff$ & $ W_nUseful_i - W_nUseful_{i-1} $ \\
$W_nUsefulDiffRatio$ & $ \frac{W_nUseful_{i} - W_nUseful_{i-1}}{W_nUseful_{i-1}} $ \\
$W_nUsefulAverage$ & $ \frac{W_nUseful_i}{W_nReview_i} $ \\
$W_nUsefulAverageDiff$ & $ W_nUsefulAverage_{i} - W_nUsefulAverage_{i-1} $ \\
$W_nUsefulDiffH_m$ & $ W_nUseful_{i} - W_mUseful_{i-n} $ \\
$W_nUsefulAverageDiffH_m$ & $ W_nUsefulAverage_{i} - W_mUsefulAverage_{i-n} $ \\ 

$W_nUsefulR$ & $ \sum_{i=1}^N\{1\ if\ usefulVoteCount_i>0\ else\ 0\}$ \\
$W_nUsefulRDiff$ & $ W_nUsefulR_i - W_nUsefulR_{i-1} $ \\
$W_nUsefulRDiffRatio$ & $ \frac{W_nUsefulR_{i} - W_nUsefulR_{i-1}}{W_nUsefulR_{i-1}} $ \\
$W_nUsefulRRatio$ & $ \frac{W_nUsefulR_i}{W_nReview_i} $ \\
$W_nUsefulRRatioDiff$ & $ W_nUsefulRRatio_{i} - W_nUsefulRRatio_{i-1} $ \\
$W_nUsefulRDiffH_m$ & $ W_nUsefulR_{i} - W_mUsefulR_{i-n} $ \\
$W_nUsefulRRatioDiffH_m$ & $ W_nUsefulRRatio_{i} - W_mUsefulRRatio_{i-n} $ \\ 
\hline
\end{tabular}}
\end{table}

\begin{table}[!htbp]
\centering
\setlength{\belowcaptionskip}{10pt}
\caption{Useless Features}
\label{tb_features_useless}
\resizebox{\textwidth}{!}{
\begin{tabular}{ll}
\hline
Feature & Definition \\ \hline
$W_nUseless$ & $ \sum_{i=1}^NUselessVoteCount_i$ \\
$W_nUselessDiff$ & $ W_nUseless_i - W_nUseless_{i-1} $ \\
$W_nUselessDiffRatio$ & $ \frac{W_nUseless_{i} - W_nUseless_{i-1}}{W_nUseless_{i-1}} $ \\
$W_nUselessAverage$ & $ \frac{W_nUseless_i}{W_nReview_i} $ \\
$W_nUselessAverageDiff$ & $ W_nUselessAverage_{i} - W_nUselessAverage_{i-1} $ \\
$W_nUselessDiffH_m$ & $ W_nUseless_{i} - W_mUseless_{i-n} $ \\
$W_nUselessAverageDiffH_m$ & $ W_nUselessAverage_{i} - W_mUselessAverage_{i-n} $ \\ 

$W_nUselessR$ & $ \sum_{i=1}^N\{1\ if\ UselessVoteCount_i>0\ else\ 0\}$ \\
$W_nUselessRDiff$ & $ W_nUselessR_i - W_nUselessR_{i-1} $ \\
$W_nUselessRDiffRatio$ & $ \frac{W_nUselessR_{i} - W_nUselessR_{i-1}}{W_nUselessR_{i-1}} $ \\
$W_nUselessRRatio$ & $ \frac{W_nUselessR_i}{W_nReview_i} $ \\
$W_nUselessRRatioDiff$ & $ W_nUselessRRatio_{i} - W_nUselessRRatio_{i-1} $ \\
$W_nUselessRDiffH_m$ & $ W_nUselessR_{i} - W_mUselessR_{i-n} $ \\
$W_nUselessRRatioDiffH_m$ & $ W_nUselessRRatio_{i} - W_mUselessRRatio_{i-n} $ \\ 
\hline
\end{tabular}}
\end{table}

\begin{table}[!htbp]
\centering
\setlength{\belowcaptionskip}{10pt}
\caption{Image Features}
\label{tb_features_image}
\resizebox{\textwidth}{!}{
\begin{tabular}{ll}
\hline
Feature & Definition \\ \hline
$W_nImage$ & $ \sum_{i=1}^Nimg_i$ \\
$W_nImageDiff$ & $ W_nImage_i - W_nImage_{i-1} $ \\
$W_nImageDiffRatio$ & $ \frac{W_nImage_{i} - W_nImage_{i-1}}{W_nImage_{i-1}} $ \\
$W_nImageAverage$ & $ \frac{W_nImage_i}{W_nReview_i} $ \\
$W_nImageAverageDiff$ & $ W_nImageAverage_{i} - W_nImageAverage_{i-1} $ \\
$W_nImageDiffH_m$ & $ W_nImage_{i} - W_mImage_{i-n} $ \\
$W_nImageAverageDiffH_m$ & $ W_nImageAverage_{i} - W_mImageAverage_{i-n} $ \\ 

$W_nImageR$ & $ \sum_{i=1}^N\{1\ if\ img_i>0\ else\ 0\}$ \\
$W_nImageRDiff$ & $ W_nImageR_i - W_nImageR_{i-1} $ \\
$W_nImageRDiffRatio$ & $ \frac{W_nImageR_{i} - W_nImageR_{i-1}}{W_nImageR_{i-1}} $ \\
$W_nImageRRatio$ & $ \frac{W_nImageR_i}{W_nReview_i} $ \\
$W_nImageRRatioDiff$ & $ W_nImageRRatio_{i} - W_nImageRRatio_{i-1} $ \\
$W_nImageRDiffH_m$ & $ W_nImageR_{i} - W_mImageR_{i-n} $ \\
$W_nImageRRatioDiffH_m$ & $ W_nImageRRatio_{i} - W_mImageRRatio_{i-n} $ \\ 
\hline
\end{tabular}}
\end{table}

\begin{table}[!htbp]
\centering
\setlength{\belowcaptionskip}{10pt}
\caption{Reply Features}
\label{tb_features_reply}
\resizebox{\textwidth}{!}{
\begin{tabular}{ll}
\hline
Feature & Definition \\ \hline
$W_nReply$ & $ \sum_{i=1}^NreplyCount_i$ \\
$W_nReplyDiff$ & $ W_nReply_i - W_nReply_{i-1} $ \\
$W_nReplyDiffRatio$ & $ \frac{W_nReply_{i} - W_nReply_{i-1}}{W_nReply_{i-1}} $ \\
$W_nReplyAverage$ & $ \frac{W_nReply_i}{W_nReview_i} $ \\
$W_nReplyAverageDiff$ & $ W_nReplyAverage_{i} - W_nReplyAverage_{i-1} $ \\
$W_nReplyDiffH_m$ & $ W_nReply_{i} - W_mReply_{i-n} $ \\
$W_nReplyAverageDiffH_m$ & $ W_nReplyAverage_{i} - W_mReplyAverage_{i-n} $ \\ 

$W_nReplyR$ & $ \sum_{i=1}^N\{1\ if\ replyCount_i>0\ else\ 0\}$ \\
$W_nReplyRDiff$ & $ W_nReplyR_i - W_nReplyR_{i-1} $ \\
$W_nReplyRDiffRatio$ & $ \frac{W_nReplyR_{i} - W_nReplyR_{i-1}}{W_nReplyR_{i-1}} $ \\
$W_nReplyRRatio$ & $ \frac{W_nReplyR_i}{W_nReview_i} $ \\
$W_nReplyRRatioDiff$ & $ W_nReplyRRatio_{i} - W_nReplyRRatio_{i-1} $ \\
$W_nReplyRDiffH_m$ & $ W_nReplyR_{i} - W_mReplyR_{i-n} $ \\
$W_nReplyRRatioDiffH_m$ & $ W_nReplyRRatio_{i} - W_mImageRRatio_{i-n} $ \\ 
\hline
\end{tabular}}
\end{table}

\begin{table}[!htbp]
\centering
\setlength{\belowcaptionskip}{10pt}
\caption{Client Features}
\label{tb_features_client}
\resizebox{\textwidth}{!}{
\begin{tabular}{ll}
\hline
Feature & Definition \\ \hline
$W_nClient^c$ & $ \sum_{i=1}^N\{1\ if\ userClient_i==c\ else\ 0 \}$ \\
$W_nClient^cDiff$ & $ W_nClient^c_i - W_nClient^c_{i-1} $ \\
$W_nClient^cDiffRatio$ & $ \frac{W_nClient^c_{i} - W_nClient^c_{i-1}}{W_nClient^c_{i-1}} $ \\
$W_nClient^cRatio$ & $ \frac{W_nClient^c_i}{W_nReview_i} $ \\
$W_nClient^cRatioDiff$ & $ W_nClient^cRatio_{i} - W_nClient^cRatio_{i-1} $ \\
$W_nClient^cDiffH_m$ & $ W_nClient^c_{i} - W_mClient^c_{i-n} $ \\
$W_nClient^cRatioDiffH_m$ & $ W_nClient^cRatio_{i} - W_mClient^cRatio_{i-n} $ \\ 
\hline
\end{tabular}}
\end{table}

\begin{table}[!htbp]
\centering
\setlength{\belowcaptionskip}{10pt}
\caption{Mobile Features}
\label{tb_features_mobile}
\resizebox{0.8\textwidth}{!}{
\begin{tabular}{ll}
\hline
Feature & Definition \\ \hline
$W_nMobile$ & $ \sum_{i=1}^N\{1\ if\ isMobile_i==True\ else\ 0\}$ \\
$W_nMobileDiff$ & $ W_nMobile_i - W_nMobile_{i-1} $ \\
$W_nMobileDiffRatio$ & $ \frac{W_nMobile_{i} - W_nMobile_{i-1}}{W_nMobile_{i-1}} $ \\
$W_nMobileRatio$ & $ \frac{W_nMobile_i}{W_nReview_i} $ \\
$W_nMobileRatioDiff$ & $ W_nMobileRatio_{i} - W_nMobileRatio_{i-1} $ \\
$W_nMobileDiffH_m$ & $ W_nMobile_{i} - W_mMobile_{i-n} $ \\
$W_nMobileRatioDiffH_m$ & $ W_nMobileRatio_{i} - W_mMobileRatio_{i-n} $ \\ \hline
\end{tabular}}
\end{table}

\begin{table}[!htbp]
\centering
\setlength{\belowcaptionskip}{10pt}
\caption{Ten technical indicators and their definitions}
\label{ten_indicator}
\resizebox{\textwidth}{!}{%
\begin{tabular}{ll}
\hline
Name of Indicators & Definition \\ \hline
Simple n(10 here)-week Moving Average & $\frac{C_t + C_{t-1} + ... + C_{t-9}}{n}$ \\
Weighted n(10 here)-week Moving Average & $\frac{10\times C_t +9\times C_{t-1} + ... + C_{t-9}}{n+(n-1)+...+1}$ \\
Momentum & $C_t - C_{t-9}$ \\
Stochastic K\% & $\frac{C_t - LL_{t-(n-1)}}{HH_{t-(n-1)} - LL_{t-(n-1)}}\times 100$ \\
Stochastic D\% & $\frac{\sum_{i=0}^{n-1}K_{t-1}}{10}\%$ \\
Relative Strength Index(RSI) & $100 - \frac{100}{1+(\sum_{i=0}^{n-1}UP_{t-i}/n)/(\sum_{i=0}^{n-1}DW_{t-i}/n)}$ \\
Moving Average Convergence Divergence(MACD) & $MACD(n)_{t-1}+\frac{2}{n+1}\times (DIFF_t-MACD(n)_{t-1})$ \\
Larry William's R\% & $\frac{H_n-C_t}{H_n-L_n}\times 100$ \\
A/D (Accumulation/Distribution) Oscillator & $\frac{H_t-C_{t-1}}{H_t-L_t}$ \\
CCI (Commodity Channel Index) & $\frac{M_t-SM_t}{0.015D_t}$ \\ \hline
\end{tabular}}
\footnotesize{
\begin{flushleft}
$C_t$ is the weekly close price i.e. $CLOSE_{W_t}$, $L_t$ is the weekly low price, i.e., $LOW_{W_t}$ and $H_t$ is the weekly high price, i.e., $HIGH_{W_t}$ at week $t$. $DIFF_t = EMA(12)_t-EMA(26)_t$, $EMA$ is the exponential moving average, $EMA(k)_t=EMA(k)_{t-1}+\alpha \times(C_t-EMA(k)_{t-1})$, $\alpha$ is a smoothing factor which is equal to $\frac{2}{k+1}$, $k$ is the time period of $k$-week exponential moving average. $LL_t$ and $HH_t$ implies lowest low and highest high in last $t$ weeks, respectively. $M_t = \frac{H_t+L_t+C_t}{3}$, $SM_t=\frac{\sum_{i=1}^nM_{t-i+1}}{n}$, $D_t=\frac{\sum_{i=1}^n\left|{M_{t-i+1} - SM_t}\right|}{n}$. $UP_t$ means upward price change while $DW_t$ is the downward price change at week $t$.
\end{flushleft}}
\end{table}


\begin{thebibliography}{90}
\expandafter\ifx\csname natexlab\endcsname\relax\def\natexlab#1{#1}\fi
\providecommand{\url}[1]{\texttt{#1}}
\providecommand{\href}[2]{#2}
\providecommand{\path}[1]{#1}
\providecommand{\DOIprefix}{doi:}
\providecommand{\ArXivprefix}{arXiv:}
\providecommand{\URLprefix}{URL: }
\providecommand{\Pubmedprefix}{pmid:}
\providecommand{\doi}[1]{\href{http://dx.doi.org/#1}{\path{#1}}}
\providecommand{\Pubmed}[1]{\href{pmid:#1}{\path{#1}}}
\providecommand{\bibinfo}[2]{#2}
\ifx\xfnm\relax \def\xfnm[#1]{\unskip,\space#1}\fi
\bibitem[{Aksoy et~al.(2008)Aksoy, Cooil, Groening and
  Keiningham}]{aksoy2008long}
\bibinfo{author}{Aksoy, L.}, \bibinfo{author}{Cooil, B.},
  \bibinfo{author}{Groening, C.}, \bibinfo{author}{Keiningham, T.L.},
  \bibinfo{year}{2008}.
\newblock \bibinfo{title}{The long-term stock market valuation of customer
  satisfaction}.
\newblock \bibinfo{journal}{Journal of Marketing} \bibinfo{volume}{72},
  \bibinfo{pages}{105--122}.
\bibitem[{Arnould and Price(1993)}]{arnould1993river}
\bibinfo{author}{Arnould, E.J.}, \bibinfo{author}{Price, L.L.},
  \bibinfo{year}{1993}.
\newblock \bibinfo{title}{River magic: Extraordinary experience and the
  extended service encounter}.
\newblock \bibinfo{journal}{Journal of Consumer Research} \bibinfo{volume}{20},
  \bibinfo{pages}{24--45}.
\bibitem[{Bambauer-Sachse and Mangold(2011)}]{bambauer2011brand}
\bibinfo{author}{Bambauer-Sachse, S.}, \bibinfo{author}{Mangold, S.},
  \bibinfo{year}{2011}.
\newblock \bibinfo{title}{Brand equity dilution through negative online
  word-of-mouth communication}.
\newblock \bibinfo{journal}{Journal of Retailing and Consumer Services}
  \bibinfo{volume}{18}, \bibinfo{pages}{38--45}.
\bibitem[{Baralis et~al.(2017)Baralis, Cagliero, Cerquitelli, Garza and
  Pulvirenti}]{baralis2017discovering}
\bibinfo{author}{Baralis, E.}, \bibinfo{author}{Cagliero, L.},
  \bibinfo{author}{Cerquitelli, T.}, \bibinfo{author}{Garza, P.},
  \bibinfo{author}{Pulvirenti, F.}, \bibinfo{year}{2017}.
\newblock \bibinfo{title}{Discovering profitable stocks for intraday trading}.
\newblock \bibinfo{journal}{Information Sciences} \bibinfo{volume}{405},
  \bibinfo{pages}{91--106}.
\bibitem[{Belk(1988)}]{belk1988possessions}
\bibinfo{author}{Belk, R.W.}, \bibinfo{year}{1988}.
\newblock \bibinfo{title}{Possessions and the extended self}.
\newblock \bibinfo{journal}{Journal of Consumer Research} \bibinfo{volume}{15},
  \bibinfo{pages}{139--168}.
\bibitem[{Blum and Langley(1997)}]{blum1997selection}
\bibinfo{author}{Blum, A.L.}, \bibinfo{author}{Langley, P.},
  \bibinfo{year}{1997}.
\newblock \bibinfo{title}{Selection of relevant features and examples in
  machine learning}.
\newblock \bibinfo{journal}{Artificial Intelligence} \bibinfo{volume}{97},
  \bibinfo{pages}{245--271}.
\bibitem[{Cabral(2000)}]{cabral2000stretching}
\bibinfo{author}{Cabral, L.M.}, \bibinfo{year}{2000}.
\newblock \bibinfo{title}{Stretching firm and brand reputation}.
\newblock \bibinfo{journal}{RAND Journal of Economics} ,
  \bibinfo{pages}{658--673}.
\bibitem[{Chen and Chen(2015)}]{chen2015hybrid}
\bibinfo{author}{Chen, M.Y.}, \bibinfo{author}{Chen, B.T.},
  \bibinfo{year}{2015}.
\newblock \bibinfo{title}{A hybrid fuzzy time series model based on granular
  computing for stock price forecasting}.
\newblock \bibinfo{journal}{Information Sciences} \bibinfo{volume}{294},
  \bibinfo{pages}{227--241}.
\bibitem[{Chen et~al.(1986)Chen, Roll and Ross}]{chen1986economic}
\bibinfo{author}{Chen, N.F.}, \bibinfo{author}{Roll, R.},
  \bibinfo{author}{Ross, S.A.}, \bibinfo{year}{1986}.
\newblock \bibinfo{title}{Economic forces and the stock market}.
\newblock \bibinfo{journal}{Journal of Business} , \bibinfo{pages}{383--403}.
\bibitem[{Chen et~al.(2012)Chen, Liu and Zhang}]{chen2012third}
\bibinfo{author}{Chen, Y.}, \bibinfo{author}{Liu, Y.}, \bibinfo{author}{Zhang,
  J.}, \bibinfo{year}{2012}.
\newblock \bibinfo{title}{When do third-party product reviews affect firm value
  and what can firms do? the case of media critics and professional movie
  reviews}.
\newblock \bibinfo{journal}{Journal of Marketing} \bibinfo{volume}{76},
  \bibinfo{pages}{116--134}.
\bibitem[{Chevalier and Mayzlin(2006)}]{chevalier2006effect}
\bibinfo{author}{Chevalier, J.A.}, \bibinfo{author}{Mayzlin, D.},
  \bibinfo{year}{2006}.
\newblock \bibinfo{title}{The effect of word of mouth on sales: Online book
  reviews}.
\newblock \bibinfo{journal}{Journal of Marketing Research}
  \bibinfo{volume}{43}, \bibinfo{pages}{345--354}.
\bibitem[{Clemons et~al.(2006)Clemons, Gao and Hitt}]{clemons2006online}
\bibinfo{author}{Clemons, E.K.}, \bibinfo{author}{Gao, G.G.},
  \bibinfo{author}{Hitt, L.M.}, \bibinfo{year}{2006}.
\newblock \bibinfo{title}{When online reviews meet hyperdifferentiation: A
  study of the craft beer industry}.
\newblock \bibinfo{journal}{Journal of Management Information Systems}
  \bibinfo{volume}{23}, \bibinfo{pages}{149--171}.
\bibitem[{Creamer and Freund(2010)}]{creamer2010automated}
\bibinfo{author}{Creamer, G.}, \bibinfo{author}{Freund, Y.},
  \bibinfo{year}{2010}.
\newblock \bibinfo{title}{Automated trading with boosting and expert
  weighting}.
\newblock \bibinfo{journal}{Quantitative Finance} \bibinfo{volume}{10},
  \bibinfo{pages}{401--420}.
\bibitem[{De~Bondt and Thaler(1985)}]{de1985does}
\bibinfo{author}{De~Bondt, W.F.}, \bibinfo{author}{Thaler, R.},
  \bibinfo{year}{1985}.
\newblock \bibinfo{title}{Does the stock market overreact?}
\newblock \bibinfo{journal}{The Journal of Finance} \bibinfo{volume}{40},
  \bibinfo{pages}{793--805}.
\bibitem[{Ding et~al.(2014)Ding, Zhang, Liu and Duan}]{ding2014using}
\bibinfo{author}{Ding, X.}, \bibinfo{author}{Zhang, Y.}, \bibinfo{author}{Liu,
  T.}, \bibinfo{author}{Duan, J.}, \bibinfo{year}{2014}.
\newblock \bibinfo{title}{Using structured events to predict stock price
  movement: An empirical investigation}, in: \bibinfo{booktitle}{Proceedings of
  the 2014 Conference on Empirical Methods in Natural Language Processing
  (EMNLP)}, pp. \bibinfo{pages}{1415--1425}.
\bibitem[{Ding et~al.(2015)Ding, Zhang, Liu and Duan}]{ding2015deep}
\bibinfo{author}{Ding, X.}, \bibinfo{author}{Zhang, Y.}, \bibinfo{author}{Liu,
  T.}, \bibinfo{author}{Duan, J.}, \bibinfo{year}{2015}.
\newblock \bibinfo{title}{Deep learning for event-driven stock prediction}, in:
  \bibinfo{booktitle}{Twenty-Fourth International Joint Conference on
  Artificial Intelligence}.
\bibitem[{Duan et~al.(2013)Duan, Liu and Zeng}]{duan2013posterior}
\bibinfo{author}{Duan, J.}, \bibinfo{author}{Liu, H.}, \bibinfo{author}{Zeng,
  J.}, \bibinfo{year}{2013}.
\newblock \bibinfo{title}{Posterior probability model for stock return
  prediction based on analyst’s recommendation behavior}.
\newblock \bibinfo{journal}{Knowledge-Based Systems} \bibinfo{volume}{50},
  \bibinfo{pages}{151--158}.
\bibitem[{Efendi et~al.(2018)Efendi, Arbaiy and Deris}]{efendi2018new}
\bibinfo{author}{Efendi, R.}, \bibinfo{author}{Arbaiy, N.},
  \bibinfo{author}{Deris, M.M.}, \bibinfo{year}{2018}.
\newblock \bibinfo{title}{A new procedure in stock market forecasting based on
  fuzzy random auto-regression time series model}.
\newblock \bibinfo{journal}{Information Sciences} \bibinfo{volume}{441},
  \bibinfo{pages}{113--132}.
\bibitem[{Engelberg and Gao(2011)}]{engelberg2011search}
\bibinfo{author}{Engelberg, J.}, \bibinfo{author}{Gao, P.},
  \bibinfo{year}{2011}.
\newblock \bibinfo{title}{In search of attention}.
\newblock \bibinfo{journal}{The Journal of Finance} \bibinfo{volume}{66},
  \bibinfo{pages}{1461--1499}.
\bibitem[{Faircloth et~al.(2001)Faircloth, Capella and
  Alford}]{faircloth2001effect}
\bibinfo{author}{Faircloth, J.B.}, \bibinfo{author}{Capella, L.M.},
  \bibinfo{author}{Alford, B.L.}, \bibinfo{year}{2001}.
\newblock \bibinfo{title}{The effect of brand attitude and brand image on brand
  equity}.
\newblock \bibinfo{journal}{Journal of Marketing Theory and Practice}
  \bibinfo{volume}{9}, \bibinfo{pages}{61--75}.
\bibitem[{Fornell et~al.(2016)Fornell, Morgeson~III and
  Hult}]{fornell2016stock}
\bibinfo{author}{Fornell, C.}, \bibinfo{author}{Morgeson~III, F.V.},
  \bibinfo{author}{Hult, G.T.M.}, \bibinfo{year}{2016}.
\newblock \bibinfo{title}{Stock returns on customer satisfaction do beat the
  market: gauging the effect of a marketing intangible}.
\newblock \bibinfo{journal}{Journal of Marketing} \bibinfo{volume}{80},
  \bibinfo{pages}{92--107}.
\bibitem[{Freund et~al.(1996)Freund, Schapire et~al.}]{freund1996experiments}
\bibinfo{author}{Freund, Y.}, \bibinfo{author}{Schapire, R.E.}, et~al.,
  \bibinfo{year}{1996}.
\newblock \bibinfo{title}{Experiments with a new boosting algorithm}, in:
  \bibinfo{booktitle}{icml}, \bibinfo{organization}{Citeseer}. pp.
  \bibinfo{pages}{148--156}.
\bibitem[{Friedman et~al.(2000)Friedman, Hastie, Tibshirani
  et~al.}]{friedman2000additive}
\bibinfo{author}{Friedman, J.}, \bibinfo{author}{Hastie, T.},
  \bibinfo{author}{Tibshirani, R.}, et~al., \bibinfo{year}{2000}.
\newblock \bibinfo{title}{Additive logistic regression: a statistical view of
  boosting (with discussion and a rejoinder by the authors)}.
\newblock \bibinfo{journal}{The Annals of Statistics} \bibinfo{volume}{28},
  \bibinfo{pages}{337--407}.
\bibitem[{Geva and Zahavi(2014)}]{geva2014empirical}
\bibinfo{author}{Geva, T.}, \bibinfo{author}{Zahavi, J.}, \bibinfo{year}{2014}.
\newblock \bibinfo{title}{Empirical evaluation of an automated intraday stock
  recommendation system incorporating both market data and textual news}.
\newblock \bibinfo{journal}{Decision Support Systems} \bibinfo{volume}{57},
  \bibinfo{pages}{212--223}.
\bibitem[{Godes and Mayzlin(2009)}]{godes2009firm}
\bibinfo{author}{Godes, D.}, \bibinfo{author}{Mayzlin, D.},
  \bibinfo{year}{2009}.
\newblock \bibinfo{title}{Firm-created word-of-mouth communication: Evidence
  from a field test}.
\newblock \bibinfo{journal}{Marketing Science} \bibinfo{volume}{28},
  \bibinfo{pages}{721--739}.
\bibitem[{Grace and O'Cass(2004)}]{grace2004examining}
\bibinfo{author}{Grace, D.}, \bibinfo{author}{O'Cass, A.},
  \bibinfo{year}{2004}.
\newblock \bibinfo{title}{Examining service experiences and post-consumption
  evaluations}.
\newblock \bibinfo{journal}{Journal of Services Marketing}
  \bibinfo{volume}{18}, \bibinfo{pages}{450--461}.
\bibitem[{Göçken et~al.(2016)Göçken, Özçalıcı, Boru and
  Dosdoğru}]{goccken2016integrating}
\bibinfo{author}{Göçken, M.}, \bibinfo{author}{Özçalıcı, M.},
  \bibinfo{author}{Boru, A.}, \bibinfo{author}{Dosdoğru, A.T.},
  \bibinfo{year}{2016}.
\newblock \bibinfo{title}{Integrating metaheuristics and artificial neural
  networks for improved stock price prediction}.
\newblock \bibinfo{journal}{Expert Systems with Applications}
  \bibinfo{volume}{44}, \bibinfo{pages}{320--331}.
\bibitem[{Hagenau et~al.(2013)Hagenau, Liebmann and
  Neumann}]{hagenau2013automated}
\bibinfo{author}{Hagenau, M.}, \bibinfo{author}{Liebmann, M.},
  \bibinfo{author}{Neumann, D.}, \bibinfo{year}{2013}.
\newblock \bibinfo{title}{Automated news reading: Stock price prediction based
  on financial news using context-capturing features}.
\newblock \bibinfo{journal}{Decision Support Systems} \bibinfo{volume}{55},
  \bibinfo{pages}{685--697}.
\bibitem[{Harris(1986)}]{harris1986transaction}
\bibinfo{author}{Harris, L.}, \bibinfo{year}{1986}.
\newblock \bibinfo{title}{A transaction data study of weekly and intradaily
  patterns in stock returns}.
\newblock \bibinfo{journal}{Journal of Financial Economics}
  \bibinfo{volume}{16}, \bibinfo{pages}{99--117}.
\bibitem[{Hoch(2002)}]{hoch2002product}
\bibinfo{author}{Hoch, S.J.}, \bibinfo{year}{2002}.
\newblock \bibinfo{title}{Product experience is seductive}.
\newblock \bibinfo{journal}{Journal of Consumer Research} \bibinfo{volume}{29},
  \bibinfo{pages}{448--454}.
\bibitem[{Hoch and Deighton(1989)}]{hoch1989managing}
\bibinfo{author}{Hoch, S.J.}, \bibinfo{author}{Deighton, J.},
  \bibinfo{year}{1989}.
\newblock \bibinfo{title}{Managing what consumers learn from experience}.
\newblock \bibinfo{journal}{Journal of Marketing} \bibinfo{volume}{53},
  \bibinfo{pages}{1--20}.
\bibitem[{Holbrook(2000)}]{holbrook2000millennial}
\bibinfo{author}{Holbrook, M.B.}, \bibinfo{year}{2000}.
\newblock \bibinfo{title}{The millennial consumer in the texts of our times:
  Experience and entertainment}.
\newblock \bibinfo{journal}{Journal of Macromarketing} \bibinfo{volume}{20},
  \bibinfo{pages}{178--192}.
\bibitem[{Holmes et~al.(2013)Holmes, Byrne and Rowley}]{holmes2013mobile}
\bibinfo{author}{Holmes, A.}, \bibinfo{author}{Byrne, A.},
  \bibinfo{author}{Rowley, J.}, \bibinfo{year}{2013}.
\newblock \bibinfo{title}{Mobile shopping behaviour: insights into attitudes,
  shopping process involvement and location}.
\newblock \bibinfo{journal}{International Journal of Retail \& Distribution
  Management} \bibinfo{volume}{42}, \bibinfo{pages}{25--39}.
\bibitem[{Hsu(2011)}]{hsu2011hybrid}
\bibinfo{author}{Hsu, C.M.}, \bibinfo{year}{2011}.
\newblock \bibinfo{title}{A hybrid procedure for stock price prediction by
  integrating self-organizing map and genetic programming}.
\newblock \bibinfo{journal}{Expert Systems with Applications}
  \bibinfo{volume}{38}, \bibinfo{pages}{14026--14036}.
\bibitem[{Hu et~al.(2009)Hu, Pavlou and Zhang}]{hu2009identifying}
\bibinfo{author}{Hu, N.}, \bibinfo{author}{Pavlou, P.}, \bibinfo{author}{Zhang,
  J.}, \bibinfo{year}{2009}.
\newblock \bibinfo{title}{Identifying and overcoming self-selection biases in
  online product reviews}.
\newblock \bibinfo{type}{Technical Report}. Working paper, Temple University,
  Philadelphia.
\bibitem[{Huang(2018)}]{huang2018customer}
\bibinfo{author}{Huang, J.}, \bibinfo{year}{2018}.
\newblock \bibinfo{title}{The customer knows best: The investment value of
  consumer opinions}.
\newblock \bibinfo{journal}{Journal of Financial Economics}
  \bibinfo{volume}{128}, \bibinfo{pages}{164--182}.
\bibitem[{Huffman and Houston(1993)}]{huffman1993goal}
\bibinfo{author}{Huffman, C.}, \bibinfo{author}{Houston, M.J.},
  \bibinfo{year}{1993}.
\newblock \bibinfo{title}{Goal-oriented experiences and the development of
  knowledge}.
\newblock \bibinfo{journal}{Journal of Consumer Research} \bibinfo{volume}{20},
  \bibinfo{pages}{190--207}.
\bibitem[{Jain and Joh(1988)}]{jain1988dependence}
\bibinfo{author}{Jain, P.C.}, \bibinfo{author}{Joh, G.H.},
  \bibinfo{year}{1988}.
\newblock \bibinfo{title}{The dependence between hourly prices and trading
  volume}.
\newblock \bibinfo{journal}{Journal of Financial and Quantitative Analysis}
  \bibinfo{volume}{23}, \bibinfo{pages}{269--283}.
\bibitem[{Jasemi et~al.(2011)Jasemi, Kimiagari and
  Memariani}]{jasemi2011modern}
\bibinfo{author}{Jasemi, M.}, \bibinfo{author}{Kimiagari, A.M.},
  \bibinfo{author}{Memariani, A.}, \bibinfo{year}{2011}.
\newblock \bibinfo{title}{A modern neural network model to do stock market
  timing on the basis of the ancient investment technique of japanese
  candlestick}.
\newblock \bibinfo{journal}{Expert Systems with Applications}
  \bibinfo{volume}{38}, \bibinfo{pages}{3884--3890}.
\bibitem[{Jegadeesh and Titman(1993)}]{jegadeesh1993returns}
\bibinfo{author}{Jegadeesh, N.}, \bibinfo{author}{Titman, S.},
  \bibinfo{year}{1993}.
\newblock \bibinfo{title}{Returns to buying winners and selling losers:
  Implications for stock market efficiency}.
\newblock \bibinfo{journal}{The Journal of Finance} \bibinfo{volume}{48},
  \bibinfo{pages}{65--91}.
\bibitem[{Jones and Litzenberger(1970)}]{jones1970quarterly}
\bibinfo{author}{Jones, C.P.}, \bibinfo{author}{Litzenberger, R.H.},
  \bibinfo{year}{1970}.
\newblock \bibinfo{title}{Quarterly earnings reports and intermediate stock
  price trends}.
\newblock \bibinfo{journal}{The Journal of Finance} \bibinfo{volume}{25},
  \bibinfo{pages}{143--148}.
\bibitem[{Kim(2003)}]{kim2003financial}
\bibinfo{author}{Kim, K.j.}, \bibinfo{year}{2003}.
\newblock \bibinfo{title}{Financial time series forecasting using support
  vector machines}.
\newblock \bibinfo{journal}{Neurocomputing} \bibinfo{volume}{55},
  \bibinfo{pages}{307--319}.
\bibitem[{Kim and Ahn(2012)}]{kim2012simultaneous}
\bibinfo{author}{Kim, K.J.}, \bibinfo{author}{Ahn, H.}, \bibinfo{year}{2012}.
\newblock \bibinfo{title}{Simultaneous optimization of artificial neural
  networks for financial forecasting}.
\newblock \bibinfo{journal}{Applied Intelligence} \bibinfo{volume}{36},
  \bibinfo{pages}{887--898}.
\bibitem[{Li et~al.(2016)Li, Leng, Yang and Yu}]{li2016stock}
\bibinfo{author}{Li, L.}, \bibinfo{author}{Leng, S.}, \bibinfo{author}{Yang,
  J.}, \bibinfo{author}{Yu, M.}, \bibinfo{year}{2016}.
\newblock \bibinfo{title}{Stock market autoregressive dynamics: A multinational
  comparative study with quantile regression}.
\newblock \bibinfo{journal}{Mathematical Problems in Engineering}
  \bibinfo{volume}{2016}.
\bibitem[{Li et~al.(2014)Li, Wang, Li, Liu, Gong and Chen}]{li2014effect}
\bibinfo{author}{Li, Q.}, \bibinfo{author}{Wang, T.}, \bibinfo{author}{Li, P.},
  \bibinfo{author}{Liu, L.}, \bibinfo{author}{Gong, Q.}, \bibinfo{author}{Chen,
  Y.}, \bibinfo{year}{2014}.
\newblock \bibinfo{title}{The effect of news and public mood on stock
  movements}.
\newblock \bibinfo{journal}{Information Sciences} \bibinfo{volume}{278},
  \bibinfo{pages}{826--840}.
\bibitem[{Li and Hitt(2008)}]{li2008self}
\bibinfo{author}{Li, X.}, \bibinfo{author}{Hitt, L.M.}, \bibinfo{year}{2008}.
\newblock \bibinfo{title}{Self-selection and information role of online product
  reviews}.
\newblock \bibinfo{journal}{Information Systems Research} \bibinfo{volume}{19},
  \bibinfo{pages}{456--474}.
\bibitem[{Liu(2006)}]{liu2006word}
\bibinfo{author}{Liu, Y.}, \bibinfo{year}{2006}.
\newblock \bibinfo{title}{Word of mouth for movies: Its dynamics and impact on
  box office revenue}.
\newblock \bibinfo{journal}{Journal of Marketing} \bibinfo{volume}{70},
  \bibinfo{pages}{74--89}.
\bibitem[{Long et~al.(2019)Long, Lu and Cui}]{long2019deep}
\bibinfo{author}{Long, W.}, \bibinfo{author}{Lu, Z.}, \bibinfo{author}{Cui,
  L.}, \bibinfo{year}{2019}.
\newblock \bibinfo{title}{Deep learning-based feature engineering for stock
  price movement prediction}.
\newblock \bibinfo{journal}{Knowledge-Based Systems} \bibinfo{volume}{164},
  \bibinfo{pages}{163--173}.
\bibitem[{Luo(2009)}]{luo2009quantifying}
\bibinfo{author}{Luo, X.}, \bibinfo{year}{2009}.
\newblock \bibinfo{title}{Quantifying the long-term impact of negative word of
  mouth on cash flows and stock prices}.
\newblock \bibinfo{journal}{Marketing Science} \bibinfo{volume}{28},
  \bibinfo{pages}{148--165}.
\bibitem[{Luo and Zhang(2013)}]{luo2013consumer}
\bibinfo{author}{Luo, X.}, \bibinfo{author}{Zhang, J.}, \bibinfo{year}{2013}.
\newblock \bibinfo{title}{How do consumer buzz and traffic in social media
  marketing predict the value of the firm?}
\newblock \bibinfo{journal}{Journal of Management Information Systems}
  \bibinfo{volume}{30}, \bibinfo{pages}{213--238}.
\bibitem[{Luo et~al.(2013)Luo, Zhang and Duan}]{luo2013social}
\bibinfo{author}{Luo, X.}, \bibinfo{author}{Zhang, J.}, \bibinfo{author}{Duan,
  W.}, \bibinfo{year}{2013}.
\newblock \bibinfo{title}{Social media and firm equity value}.
\newblock \bibinfo{journal}{Information Systems Research} \bibinfo{volume}{24},
  \bibinfo{pages}{146--163}.
\bibitem[{Malbon(2013)}]{malbon2013taking}
\bibinfo{author}{Malbon, J.}, \bibinfo{year}{2013}.
\newblock \bibinfo{title}{Taking fake online consumer reviews seriously}.
\newblock \bibinfo{journal}{Journal of Consumer Policy} \bibinfo{volume}{36},
  \bibinfo{pages}{139--157}.
\bibitem[{Martin et~al.(2007)Martin, Barron and Norton}]{martin2007choosing}
\bibinfo{author}{Martin, J.}, \bibinfo{author}{Barron, G.},
  \bibinfo{author}{Norton, M.}, \bibinfo{year}{2007}.
\newblock \bibinfo{title}{Choosing to be uncertain: Preferences for high
  variance experiences}, in: \bibinfo{booktitle}{London Business School
  Trans-Atlantic Doctoral Conference}.
\bibitem[{Meenaghan(1995)}]{meenaghan1995role}
\bibinfo{author}{Meenaghan, T.}, \bibinfo{year}{1995}.
\newblock \bibinfo{title}{The role of advertising in brand image development}.
\newblock \bibinfo{journal}{Journal of Product \& Brand Management}
  \bibinfo{volume}{4}, \bibinfo{pages}{23--34}.
\bibitem[{Menkhoff(2010)}]{menkhoff2010use}
\bibinfo{author}{Menkhoff, L.}, \bibinfo{year}{2010}.
\newblock \bibinfo{title}{The use of technical analysis by fund managers:
  International evidence}.
\newblock \bibinfo{journal}{Journal of Banking \& Finance}
  \bibinfo{volume}{34}, \bibinfo{pages}{2573--2586}.
\bibitem[{Mittal et~al.(2005)Mittal, Anderson, Sayrak and
  Tadikamalla}]{mittal2005dual}
\bibinfo{author}{Mittal, V.}, \bibinfo{author}{Anderson, E.W.},
  \bibinfo{author}{Sayrak, A.}, \bibinfo{author}{Tadikamalla, P.},
  \bibinfo{year}{2005}.
\newblock \bibinfo{title}{Dual emphasis and the long-term financial impact of
  customer satisfaction}.
\newblock \bibinfo{journal}{Marketing Science} \bibinfo{volume}{24},
  \bibinfo{pages}{544--555}.
\bibitem[{Morgan and Rego(2006)}]{morgan2006value}
\bibinfo{author}{Morgan, N.A.}, \bibinfo{author}{Rego, L.L.},
  \bibinfo{year}{2006}.
\newblock \bibinfo{title}{The value of different customer satisfaction and
  loyalty metrics in predicting business performance}.
\newblock \bibinfo{journal}{Marketing Science} \bibinfo{volume}{25},
  \bibinfo{pages}{426--439}.
\bibitem[{Nassirtoussi et~al.(2015)Nassirtoussi, Aghabozorgi, Wah and
  Ngo}]{nassirtoussi2015text}
\bibinfo{author}{Nassirtoussi, A.K.}, \bibinfo{author}{Aghabozorgi, S.},
  \bibinfo{author}{Wah, T.Y.}, \bibinfo{author}{Ngo, D.C.L.},
  \bibinfo{year}{2015}.
\newblock \bibinfo{title}{Text mining of news-headlines for forex market
  prediction: A multi-layer dimension reduction algorithm with semantics and
  sentiment}.
\newblock \bibinfo{journal}{Expert Systems with Applications}
  \bibinfo{volume}{42}, \bibinfo{pages}{306--324}.
\bibitem[{Nayak et~al.(2015)Nayak, Mishra and Rath}]{nayak2015naive}
\bibinfo{author}{Nayak, R.K.}, \bibinfo{author}{Mishra, D.},
  \bibinfo{author}{Rath, A.K.}, \bibinfo{year}{2015}.
\newblock \bibinfo{title}{A na{\"\i}ve svm-knn based stock market trend
  reversal analysis for indian benchmark indices}.
\newblock \bibinfo{journal}{Applied Soft Computing} \bibinfo{volume}{35},
  \bibinfo{pages}{670--680}.
\bibitem[{O’Connor and Madden(2006)}]{o2006neural}
\bibinfo{author}{O’Connor, N.}, \bibinfo{author}{Madden, M.G.},
  \bibinfo{year}{2006}.
\newblock \bibinfo{title}{A neural network approach to predicting stock
  exchange movements using external factors}.
\newblock \bibinfo{journal}{Knowledge-Based Systems} \bibinfo{volume}{5},
  \bibinfo{pages}{371--378}.
\bibitem[{Pan et~al.(2017)Pan, Xiao, Wang and Yang}]{pan2017multiple}
\bibinfo{author}{Pan, Y.}, \bibinfo{author}{Xiao, Z.}, \bibinfo{author}{Wang,
  X.}, \bibinfo{author}{Yang, D.}, \bibinfo{year}{2017}.
\newblock \bibinfo{title}{A multiple support vector machine approach to stock
  index forecasting with mixed frequency sampling}.
\newblock \bibinfo{journal}{Knowledge-Based Systems} \bibinfo{volume}{122},
  \bibinfo{pages}{90--102}.
\bibitem[{Parker(2009)}]{parker2009comparison}
\bibinfo{author}{Parker, B.T.}, \bibinfo{year}{2009}.
\newblock \bibinfo{title}{A comparison of brand personality and brand
  user-imagery congruence}.
\newblock \bibinfo{journal}{Journal of Consumer Marketing}
  \bibinfo{volume}{26}, \bibinfo{pages}{175--184}.
\bibitem[{Patel et~al.(2015)Patel, Shah, Thakkar and
  Kotecha}]{patel2015predicting}
\bibinfo{author}{Patel, J.}, \bibinfo{author}{Shah, S.},
  \bibinfo{author}{Thakkar, P.}, \bibinfo{author}{Kotecha, K.},
  \bibinfo{year}{2015}.
\newblock \bibinfo{title}{Predicting stock market index using fusion of machine
  learning techniques}.
\newblock \bibinfo{journal}{Expert Systems with Applications}
  \bibinfo{volume}{42}, \bibinfo{pages}{2162--2172}.
\bibitem[{Plummer(2000)}]{plummer2000personality}
\bibinfo{author}{Plummer, J.T.}, \bibinfo{year}{2000}.
\newblock \bibinfo{title}{How personality makes a difference}.
\newblock \bibinfo{journal}{Journal of Advertising Research}
  \bibinfo{volume}{40}, \bibinfo{pages}{79--83}.
\bibitem[{Preis et~al.(2013)Preis, Moat and Stanley}]{preis2013quantifying}
\bibinfo{author}{Preis, T.}, \bibinfo{author}{Moat, H.S.},
  \bibinfo{author}{Stanley, H.E.}, \bibinfo{year}{2013}.
\newblock \bibinfo{title}{Quantifying trading behavior in financial markets
  using google trends}.
\newblock \bibinfo{journal}{Scientific Reports} \bibinfo{volume}{3},
  \bibinfo{pages}{1684}.
\bibitem[{Ruan et~al.(2018)Ruan, Durresi and Alfantoukh}]{ruan2018using}
\bibinfo{author}{Ruan, Y.}, \bibinfo{author}{Durresi, A.},
  \bibinfo{author}{Alfantoukh, L.}, \bibinfo{year}{2018}.
\newblock \bibinfo{title}{Using twitter trust network for stock market
  analysis}.
\newblock \bibinfo{journal}{Knowledge-Based Systems} \bibinfo{volume}{145},
  \bibinfo{pages}{207--218}.
\bibitem[{Schapire and Singer(1999)}]{schapire1999improved}
\bibinfo{author}{Schapire, R.E.}, \bibinfo{author}{Singer, Y.},
  \bibinfo{year}{1999}.
\newblock \bibinfo{title}{Improved boosting algorithms using confidence-rated
  predictions}.
\newblock \bibinfo{journal}{Machine Learning} \bibinfo{volume}{37},
  \bibinfo{pages}{297--336}.
\bibitem[{Senecal and Nantel(2004)}]{senecal2004influence}
\bibinfo{author}{Senecal, S.}, \bibinfo{author}{Nantel, J.},
  \bibinfo{year}{2004}.
\newblock \bibinfo{title}{The influence of online product recommendations on
  consumers’ online choices}.
\newblock \bibinfo{journal}{Journal of Retailing} \bibinfo{volume}{80},
  \bibinfo{pages}{159--169}.
\bibitem[{Sirgy(1982)}]{sirgy1982self}
\bibinfo{author}{Sirgy, M.J.}, \bibinfo{year}{1982}.
\newblock \bibinfo{title}{Self-concept in consumer behavior: A critical
  review}.
\newblock \bibinfo{journal}{Journal of Consumer Research} \bibinfo{volume}{9},
  \bibinfo{pages}{287--300}.
\bibitem[{Smith(1776)}]{smith1937wealth}
\bibinfo{author}{Smith, A.}, \bibinfo{year}{1776}.
\newblock \bibinfo{title}{The wealth of nations} .
\bibitem[{Subrahmanyam and Titman(1999)}]{subrahmanyam1999going}
\bibinfo{author}{Subrahmanyam, A.}, \bibinfo{author}{Titman, S.},
  \bibinfo{year}{1999}.
\newblock \bibinfo{title}{The going-public decision and the development of
  financial markets}.
\newblock \bibinfo{journal}{The Journal of Finance} \bibinfo{volume}{54},
  \bibinfo{pages}{1045--1082}.
\bibitem[{Sun et~al.(2017)Sun, Wang, Zhang, Cao, Liu and
  Wang}]{sun2017predicting}
\bibinfo{author}{Sun, T.}, \bibinfo{author}{Wang, J.}, \bibinfo{author}{Zhang,
  P.}, \bibinfo{author}{Cao, Y.}, \bibinfo{author}{Liu, B.},
  \bibinfo{author}{Wang, D.}, \bibinfo{year}{2017}.
\newblock \bibinfo{title}{Predicting stock price returns using microblog
  sentiment for chinese stock market}, in: \bibinfo{booktitle}{2017 3rd
  International Conference on Big Data Computing and Communications (BIGCOM)},
  \bibinfo{organization}{IEEE}. pp. \bibinfo{pages}{87--96}.
\bibitem[{Tang et~al.(2014)Tang, Alelyani and Liu}]{tang2014feature}
\bibinfo{author}{Tang, J.}, \bibinfo{author}{Alelyani, S.},
  \bibinfo{author}{Liu, H.}, \bibinfo{year}{2014}.
\newblock \bibinfo{title}{Feature selection for classification: A review}.
\newblock \bibinfo{journal}{Data Classification: Algorithms and Applications} ,
  \bibinfo{pages}{37}.
\bibitem[{Tang et~al.(2012)Tang, Gu and Whinston}]{tang2012content}
\bibinfo{author}{Tang, Q.}, \bibinfo{author}{Gu, B.},
  \bibinfo{author}{Whinston, A.B.}, \bibinfo{year}{2012}.
\newblock \bibinfo{title}{Content contribution for revenue sharing and
  reputation in social media: A dynamic structural model}.
\newblock \bibinfo{journal}{Journal of Management Information Systems}
  \bibinfo{volume}{29}, \bibinfo{pages}{41--76}.
\bibitem[{Tellis and Johnson(2007)}]{tellis2007value}
\bibinfo{author}{Tellis, G.J.}, \bibinfo{author}{Johnson, J.},
  \bibinfo{year}{2007}.
\newblock \bibinfo{title}{The value of quality}.
\newblock \bibinfo{journal}{Marketing Science} \bibinfo{volume}{26},
  \bibinfo{pages}{758--773}.
\bibitem[{Tirunillai and Tellis(2012)}]{tirunillai2012does}
\bibinfo{author}{Tirunillai, S.}, \bibinfo{author}{Tellis, G.J.},
  \bibinfo{year}{2012}.
\newblock \bibinfo{title}{Does chatter really matter? dynamics of
  user-generated content and stock performance}.
\newblock \bibinfo{journal}{Marketing Science} \bibinfo{volume}{31},
  \bibinfo{pages}{198--215}.
\bibitem[{Urde(2003)}]{urde2003core}
\bibinfo{author}{Urde, M.}, \bibinfo{year}{2003}.
\newblock \bibinfo{title}{Core value-based corporate brand building}.
\newblock \bibinfo{journal}{European Journal of Marketing}
  \bibinfo{volume}{37}, \bibinfo{pages}{1017--1040}.
\bibitem[{Wang et~al.(2019)Wang, Lu and Zhao}]{wang2019aggregating}
\bibinfo{author}{Wang, H.}, \bibinfo{author}{Lu, S.}, \bibinfo{author}{Zhao,
  J.}, \bibinfo{year}{2019}.
\newblock \bibinfo{title}{Aggregating multiple types of complex data in stock
  market prediction: A model-independent framework}.
\newblock \bibinfo{journal}{Knowledge-Based Systems} \bibinfo{volume}{164},
  \bibinfo{pages}{193--204}.
\bibitem[{Wang et~al.(2016)Wang, Malthouse and Krishnamurthi}]{wang2016mobile}
\bibinfo{author}{Wang, R.J.H.}, \bibinfo{author}{Malthouse, E.C.},
  \bibinfo{author}{Krishnamurthi, L.}, \bibinfo{year}{2016}.
\newblock \bibinfo{title}{How mobile shopping affects customer purchase
  behavior: A retailer’s perspective}, in: \bibinfo{booktitle}{Let’s Get
  Engaged! Crossing the Threshold of Marketing’s Engagement Era}.
  \bibinfo{publisher}{Springer}, pp. \bibinfo{pages}{703--704}.
\bibitem[{Wang and Hua(2014)}]{wang2014semiparametric}
\bibinfo{author}{Wang, W.Y.}, \bibinfo{author}{Hua, Z.}, \bibinfo{year}{2014}.
\newblock \bibinfo{title}{A semiparametric gaussian copula regression model for
  predicting financial risks from earnings calls}, in:
  \bibinfo{booktitle}{Proceedings of the 52nd Annual Meeting of the Association
  for Computational Linguistics (Volume 1: Long Papers)}, pp.
  \bibinfo{pages}{1155--1165}.
\bibitem[{Xu et~al.(2019)Xu, Bo, Jiang and Liu}]{xu2019does}
\bibinfo{author}{Xu, Q.}, \bibinfo{author}{Bo, Z.}, \bibinfo{author}{Jiang,
  C.}, \bibinfo{author}{Liu, Y.}, \bibinfo{year}{2019}.
\newblock \bibinfo{title}{Does google search index really help predicting stock
  market volatility? evidence from a modified mixed data sampling model on
  volatility}.
\newblock \bibinfo{journal}{Knowledge-Based Systems} \bibinfo{volume}{166},
  \bibinfo{pages}{170--185}.
\bibitem[{Ye et~al.(2016)Ye, Zhang, Zhang, Fujita and Gong}]{ye2016novel}
\bibinfo{author}{Ye, F.}, \bibinfo{author}{Zhang, L.}, \bibinfo{author}{Zhang,
  D.}, \bibinfo{author}{Fujita, H.}, \bibinfo{author}{Gong, Z.},
  \bibinfo{year}{2016}.
\newblock \bibinfo{title}{A novel forecasting method based on multi-order fuzzy
  time series and technical analysis}.
\newblock \bibinfo{journal}{Information Sciences} \bibinfo{volume}{367},
  \bibinfo{pages}{41--57}.
\bibitem[{Zhang et~al.(2016)Zhang, Zhou, Kehoe and Kilic}]{zhang2016online}
\bibinfo{author}{Zhang, D.}, \bibinfo{author}{Zhou, L.},
  \bibinfo{author}{Kehoe, J.L.}, \bibinfo{author}{Kilic, I.Y.},
  \bibinfo{year}{2016}.
\newblock \bibinfo{title}{What online reviewer behaviors really matter? effects
  of verbal and nonverbal behaviors on detection of fake online reviews}.
\newblock \bibinfo{journal}{Journal of Management Information Systems}
  \bibinfo{volume}{33}, \bibinfo{pages}{456--481}.
\bibitem[{Zhang et~al.(2017)Zhang, Aggarwal and Qi}]{zhang2017stock}
\bibinfo{author}{Zhang, L.}, \bibinfo{author}{Aggarwal, C.},
  \bibinfo{author}{Qi, G.J.}, \bibinfo{year}{2017}.
\newblock \bibinfo{title}{Stock price prediction via discovering
  multi-frequency trading patterns}, in: \bibinfo{booktitle}{Proceedings of the
  23rd ACM SIGKDD International Conference on Knowledge Discovery and Data
  Mining}, \bibinfo{organization}{ACM}. pp. \bibinfo{pages}{2141--2149}.
\bibitem[{Zhang et~al.(2018)Zhang, Zhang, Wang, Yao, Fang and
  Philip}]{zhang2018improving}
\bibinfo{author}{Zhang, X.}, \bibinfo{author}{Zhang, Y.},
  \bibinfo{author}{Wang, S.}, \bibinfo{author}{Yao, Y.}, \bibinfo{author}{Fang,
  B.}, \bibinfo{author}{Philip, S.Y.}, \bibinfo{year}{2018}.
\newblock \bibinfo{title}{Improving stock market prediction via heterogeneous
  information fusion}.
\newblock \bibinfo{journal}{Knowledge-Based Systems} \bibinfo{volume}{143},
  \bibinfo{pages}{236--247}.
\bibitem[{Zhou et~al.(2015)Zhou, Lu and Fujita}]{zhou2015performance}
\bibinfo{author}{Zhou, L.}, \bibinfo{author}{Lu, D.}, \bibinfo{author}{Fujita,
  H.}, \bibinfo{year}{2015}.
\newblock \bibinfo{title}{The performance of corporate financial distress
  prediction models with features selection guided by domain knowledge and data
  mining approaches}.
\newblock \bibinfo{journal}{Knowledge-Based Systems} \bibinfo{volume}{85},
  \bibinfo{pages}{52--61}.
\bibitem[{Zhou et~al.(2017)Zhou, Si and Fujita}]{zhou2017predicting}
\bibinfo{author}{Zhou, L.}, \bibinfo{author}{Si, Y.W.},
  \bibinfo{author}{Fujita, H.}, \bibinfo{year}{2017}.
\newblock \bibinfo{title}{Predicting the listing statuses of chinese-listed
  companies using decision trees combined with an improved filter feature
  selection method}.
\newblock \bibinfo{journal}{Knowledge-Based Systems} \bibinfo{volume}{128},
  \bibinfo{pages}{93--101}.
\bibitem[{Zhou et~al.(2018)Zhou, Xu and Zhao}]{zhou2018tales}
\bibinfo{author}{Zhou, Z.}, \bibinfo{author}{Xu, K.}, \bibinfo{author}{Zhao,
  J.}, \bibinfo{year}{2018}.
\newblock \bibinfo{title}{Tales of emotion and stock in china: volatility,
  causality and prediction}.
\newblock \bibinfo{journal}{World Wide Web} \bibinfo{volume}{21},
  \bibinfo{pages}{1093--1116}.
\bibitem[{Zhou et~al.(2016)Zhou, Zhao and Xu}]{zhou2016can}
\bibinfo{author}{Zhou, Z.}, \bibinfo{author}{Zhao, J.}, \bibinfo{author}{Xu,
  K.}, \bibinfo{year}{2016}.
\newblock \bibinfo{title}{Can online emotions predict the stock market in
  china?}, in: \bibinfo{booktitle}{International Conference on Web Information
  Systems Engineering}, \bibinfo{organization}{Springer}. pp.
  \bibinfo{pages}{328--342}.
\bibitem[{Zhu and Zhang(2010)}]{zhu2010impact}
\bibinfo{author}{Zhu, F.}, \bibinfo{author}{Zhang, X.}, \bibinfo{year}{2010}.
\newblock \bibinfo{title}{Impact of online consumer reviews on sales: The
  moderating role of product and consumer characteristics}.
\newblock \bibinfo{journal}{Journal of Marketing} \bibinfo{volume}{74},
  \bibinfo{pages}{133--148}.

\end{thebibliography}
\end{document}